\documentclass[12pt]{article}
\usepackage{amsfonts}
\usepackage{amssymb}
\usepackage{graphics,psboxit,amsmath}
\usepackage{subfigure}
\usepackage{graphicx}
\usepackage{verbatim}

\def\hybrid{\topmargin -30pt    \oddsidemargin 0pt 
        \headheight 0pt \headsep 0pt
        \textwidth 6.25in       
        \textheight 9.5in       
        \marginparwidth .875in
        \parskip 5pt plus 1pt   \jot = 1.5ex}

\hybrid

\def\baselinestretch{1.2}

\catcode`\@=11

\def\marginnote#1{}
\newcount\hour
\newcount\minute
\newtoks\amorpm
\hour=\time\divide\hour by60
\minute=\time{\multiply\hour by60 \global\advance\minute by-\hour}
\edef\standardtime{{\ifnum\hour<12 \global\amorpm={am}%
        \else\global\amorpm={pm}\advance\hour by-12 \fi
        \ifnum\hour=0 \hour=12 \fi
        \number\hour:\ifnum\minute<10 0\fi\number\minute\the\amorpm}}
\edef\militarytime{\number\hour:\ifnum\minute<10 0\fi\number\minute}

\def\draftlabel#1{{\@bsphack\if@filesw {\let\thepage\relax
   \xdef\@gtempa{\write\@auxout{\string
      \newlabel{#1}{{\@currentlabel}{\thepage}}}}}\@gtempa
   \if@nobreak \ifvmode\nobreak\fi\fi\fi\@esphack}
        \gdef\@eqnlabel{#1}}
\def\@eqnlabel{}
\def\@vacuum{}
\def\draftmarginnote#1{\marginpar{\raggedright\scriptsize\tt#1}}

\def\draft{\oddsidemargin -.5truein
        \def\@oddfoot{\sl preliminary draft \hfil
        \rm\thepage\hfil\sl\today\quad\militarytime}
        \let\@evenfoot\@oddfoot \overfullrule 3pt
        \let\label=\draftlabel
        \let\marginnote=\draftmarginnote
   \def\@eqnnum{(\theequation)\rlap{\kern\marginparsep\tt\@eqnlabel}%
\global\let\@eqnlabel\@vacuum}  }

\def\draft2{
        \def\@oddfoot{\sl preliminary draft \hfil
        \rm\thepage\hfil\sl\today\quad\militarytime}
        \let\@evenfoot\@oddfoot \overfullrule 3pt
        \let\label=\draftlabel
        \let\marginnote=\draftmarginnote
   \def\@eqnnum{(\theequation)\rlap{\kern\marginparsep\tt\@eqnlabel}%
\global\let\@eqnlabel\@vacuum}  }

\def\preprint{\twocolumn\sloppy\flushbottom\parindent 2em
        \leftmargini 2em\leftmarginv .5em\leftmarginvi .5em
        \oddsidemargin -.5in    \evensidemargin -.5in
        \columnsep .4in \footheight 0pt
        \textwidth 10.in        \topmargin  -.4in
        \headheight 12pt \topskip .4in
        \textheight 6.9in \footskip 0pt
        \def\@oddhead{\thepage\hfil\addtocounter{page}{1}\thepage}
        \let\@evenhead\@oddhead \def\@oddfoot{} \def\@evenfoot{} }

\def\numberbysection{\@addtoreset{equation}{section}
        \def\theequation{\thesection.\arabic{equation}}}

\def\underline#1{\relax\ifmmode\@@underline#1\else
        $\@@underline{\hbox{#1}}$\relax\fi}

\def\titlepage{\@restonecolfalse\if@twocolumn\@restonecoltrue\onecolumn
     \else \newpage \fi \thispagestyle{empty}\c@page\z@
        \def\thefootnote{\fnsymbol{footnote}} }

\def\endtitlepage{\if@restonecol\twocolumn \else \newpage \fi
        \def\thefootnote{\arabic{footnote}}
        \setcounter{footnote}{0}}  

\catcode`@=12
\relax

\def\figcap{\section*{Figure Captions\markboth
        {FIGURECAPTIONS}{FIGURECAPTIONS}}\list
        {Figure \arabic{enumi}:\hfill}{\settowidth\labelwidth{Figure
999:}
        \leftmargin\labelwidth
        \advance\leftmargin\labelsep\usecounter{enumi}}}
 \relax
\def\tablecap{\section*{Table Captions\markboth
        {TABLECAPTIONS}{TABLECAPTIONS}}\list
        {Table \arabic{enumi}:\hfill}{\settowidth\labelwidth{Table
999:}
        \leftmargin\labelwidth
        \advance\leftmargin\labelsep\usecounter{enumi}}}
 \relax
\def\reflist{\section*{References\markboth
        {REFLIST}{REFLIST}}\list
        {[\arabic{enumi}]\hfill}{\settowidth\labelwidth{[999]}
        \leftmargin\labelwidth
        \advance\leftmargin\labelsep\usecounter{enumi}}}
 \relax
\makeatletter
\newcounter{pubctr}
\def\publist{\@ifnextchar[{\@publist}{\@@publist}}
\def\@publist[#1]{\list
        {[\arabic{pubctr}]\hfill}{\settowidth\labelwidth{[999]}
        \leftmargin\labelwidth
        \advance\leftmargin\labelsep
        \@nmbrlisttrue\def\@listctr{pubctr}
        \setcounter{pubctr}{#1}\addtocounter{pubctr}{-1}}}
\def\@@publist{\list
        {[\arabic{pubctr}]\hfill}{\settowidth\labelwidth{[999]}
        \leftmargin\labelwidth
        \advance\leftmargin\labelsep
        \@nmbrlisttrue\def\@listctr{pubctr}}}
 \relax
\makeatother

\def\be{\begin{equation}}
\def\ee{\end{equation}}
\def\ba{\begin{eqnarray}}
\def\ea{\end{eqnarray}}

\def\del{\partial}

\def\a{\alpha}

\def\b{\beta}

\def\g{\gamma}
\def\G{\Gamma}
\def\d{\delta}
\def\D{\Delta}

\def\P{\Pi}

\def\th{\theta}
\def\Th{\Theta}
\def\m{\mu}
\def\n{\nu}

\def\Om{\Omega}
\def\l{\lambda}
\def\L{\Lambda}
\def\s{\sigma}
\def\S{\Sigma}
\def\vphi{\varphi}
\def\cA{{\cal A}}
\def\cB{{\cal B}}

\def\cN{{\cal N}}

\def\cG{{\cal{G}}}
\def\cH{{\cal{H}}}

\def\cP{{\cal P}}
\def\cQ{{\cal{Q}}}
\def\cR{{\cal R}}

\def\ty{\tilde{y}}
\def\cth{{\rm c}_\th}
\def\sth{{\rm s}_\th}
\def\cpsi{{\rm c}_{\psi}}
\def\spsi{{\rm s}_{\psi}}
\def\calpha{{\rm c}_\alpha}
\def\salpha{{\rm s}_\alpha}
\def\hbeta{\hat\beta}
\def\hgamma{\hat\gamma}
\def\hsigma{\hat\sigma}

\def\elK{{\bf K}}
\def\elPi{{\bf \Pi}}
\def\elE{{\bf E}}

\def\no{\noindent}

\def\qq{\qquad}

\def\IR{\relax{\rm I\kern-.18em R}}

\def \tA  { {\tilde {A }}}

\def \ha {{1\over 2}}

\def \ov {\over}

\def\diag{{\rm diag}}
\def\const{{\rm const.}}

\def\II{\relax{\rm 1\kern-.35em1}}
\def\IR{\relax{\rm I\kern-.18em R}}
\def\inv{^{\raise.15ex\hbox{${\scriptscriptstyle -}$}\kern-.05em 1}}

\def\cR{{\cal R}}
\def\cA{{\cal A}}

\begin{document}


\renewcommand{\theequation}{\thesection.\arabic{equation}}
\csname @addtoreset\endcsname{equation}{section}

\newcommand{\beq}{\begin{equation}}
\newcommand{\eeq}[1]{\label{#1}\end{equation}}
\newcommand{\ber}{\begin{eqnarray}}
\newcommand{\eer}[1]{\label{#1}\end{eqnarray}}
\newcommand{\eqn}[1]{(\ref{#1})}
\begin{titlepage}
\begin{center}

\hfill 0704.2067 [hep-th]\\

\vskip .5in

{\large \bf Complex marginal deformations of D3-brane geometries,\\
their Penrose limits and giant gravitons}

\vskip 0.5in

{\bf Spyros D. Avramis$^{1,2}$},\phantom{x} {\bf Konstadinos
Sfetsos}$^1$\phantom{x} and\phantom{x} {\bf Dimitrios Zoakos}$^1$
\vskip 0.1in

${}^1\!$
Department of Engineering Sciences, University of Patras,\\
26110 Patras, Greece\\

\vskip .1in

${}^2\!$
Department of Physics, National Technical University of Athens,\\
15773, Athens, Greece\\

\vskip .15in

{\footnotesize {\tt avramis@mail.cern.ch}, \ \ {\tt sfetsos@upatras.gr}, \ \ {\tt dzoakos@upatras.gr}}\\

\end{center}

\vskip .4in

\centerline{\bf Abstract} \no We apply the Lunin--Maldacena
construction of gravity duals to $\b$--deformed gauge theories to
a class of Type IIB backgrounds with $U(1)^3$ global symmetry,
which include the multicenter D3-brane backgrounds dual to the
Coulomb branch of $\cN=4$ super Yang-Mills and the rotating
D3-brane backgrounds dual to the theory at finite temperature and
chemical potential. After a general discussion, we present the
full form of the deformed metrics for three special cases, which
can be used for the study of various aspects of the
marginally-deformed gauge theories. We also construct the Penrose
limits of the solutions dual to the Coulomb branch along a certain
set of geodesics and, for the resulting PP--wave metrics, we
examine the effect of $\b$--deformations on the giant graviton
states. We find that giant gravitons exist only up to a critical
value of the $\s$--deformation parameter, are not degenerate in
energy with the point graviton, and remain perturbatively stable.
Finally, we probe the $\s$--deformed multicenter solutions by
examining the static heavy-quark potential by means of Wilson
loops. We find situations that give rise to complete screening as
well as linear confinement, with the latter arising is an
intriguing way reminiscent of phase transitions in statistical
systems.


\end{titlepage}
\vfill
\eject


\tableofcontents

\newpage
\def\baselinestretch{1.2}
\baselineskip 20 pt
\no

\section{Introduction}
\label{sec1}

The AdS/CFT correspondence \cite{adscft} has proven to be an
invaluable tool for exploring the dynamics of large $N$ gauge
theories at strong coupling. In its original form, it relates
$\cN=4$, $SU(N)$ super Yang-Mills theory to Type IIB string theory
on ${\rm AdS}_5 \times {\rm S}^5$, with the limit of large 't
Hooft coupling in the gauge theory corresponding to the classical
supergravity limit of the string theory. The AdS/CFT
correspondence can be extended to less symmetric theories, a class
of which are the exactly marginal deformations of $\cN=4$ SYM,
introduced by Leigh and Strassler \cite{LS}, which break
supersymmetry down to $\cN=1$. The gravity duals of such
deformations have been identified by Lunin and Maldacena \cite{LM}
and are constructed by applying an $SL(3,\mathbb{R})$
transformation or, equivalently \cite{frolov}, a certain sequence
of T--dualities, S--dualities and coordinate shifts to the initial
${\rm AdS}_5 \times {\rm S}^5$ solution. This construction has
been generalized in \cite{frolov,frt,marginal-generalizations} and
extended to other backgrounds in \cite{marginal-otherbackgrounds}
while diverse aspects of the deformation have been examined in
\cite{marginal-various,freedman,berenstein,dorey,ZanonPenati}.

\no The Lunin--Maldacena construction can be carried over to the
Coulomb branch of the gauge theory. The latter is obtained by
moving away from the conformal point at the origin of moduli space
by giving nonzero {\em vevs} to the $SO(6)$ scalars. The
corresponding gravity duals are obtained by generalizing the
stacked-brane distribution to a multicenter one, thereby breaking
the $SO(6)$ isometry of the solutions \cite{trivedi}. A class of
marginal deformations ($\g$--deformations) of these solutions have
been obtained by the procedure outlined above in \cite{hsz}.
Probes of the resulting deformed geometries with Wilson loops,
according to the recipe of \cite{maldaloop}, have revealed a rich
structure of phenomena in the gauge theory, with behaviors ranging
from the standard Coulombic interaction to complete screening and
linear or logarithmic confinement, while the wave equation for the
radial modes of massless scalar excitations in the deformed
backgrounds turns out to be related to the Inozemtsev ${\rm BC}_1$
integrable system \cite{hsz}. Another class of marginal deformations
($\s$--deformations) of these solutions have been obtained in
\cite{ahn2}, where Wilson-loop calculations indicate the existence
of a linear confining potential in some cases.

\no A further step forward would be to extend this construction to
include the full set of complex $\b$--deformations and to apply it
for the most general case of non-extremal rotating D3-branes
\cite{trivedi,cy,rs} which are dual to the gauge theory at finite
temperature and R-charge chemical potentials and which include the
multicenter D3-branes dual to the Coulomb branch as a limiting
case. The construction of these deformed Type IIB backgrounds is
the main purpose of this paper. After constructing the deformed
solutions, we explore diverse aspects of these backgrounds, namely
the Penrose limits of the multicenter solutions, the giant
graviton states supported in the resulting PP--waves and, finally,
the  Wilson-loop heavy-quark potential of the dual gauge theory.
These investigations are carried out with emphasis on
$\s$--deformations, as their effect is often overlooked in the
literature although it is in many cases significant.

\no This paper is organized as follows: In section \ref{sec2} we
present in detail the $\b$--deformation procedure for the most
general Type IIB background consisting of the metric, a 4-form
potential and a dilaton and possessing at least a $U(1)^3$ global
symmetry. In section \ref{sec3} we consider a class of such
backgrounds, corresponding to rotating and multicenter D3-branes,
we specialize to three simple cases for which we present the
explicit form of the deformed metrics, and we demonstrate the
expected equivalence of the thermodynamics of the deformed and
undeformed metrics. In section \ref{sec4} we construct the
PP--wave backgrounds arising as Penrose limits of the deformed
multicenter solutions along a certain set of BPS and non-BPS
geodesics. In section \ref{sec5}, we consider the simplest
PP--wave background of this type and we investigate the effect of
$\s$--deformations on the energetics of giant gravitons supported
by this geometry. In section \ref{sec6} we probe the deformed
multicenter geometries by static Wilson loops for the case of
$\s$--deformations. Finally, in section \ref{sec7} we summarize
and conclude. Our conventions for T-- and S--duality are
summarized in the appendix.

\section{Marginal Deformations of Type IIB backgrounds with $U(1)^3$ isometry}
\label{sec2}

\subsection{Marginally deformed $\cN=4$ SYM and its gravity dual}

On the gauge-theory side, our general setup refers to a class of
exactly marginal deformations of $\cN=4$ super Yang-Mills theory,
namely the Leigh--Strassler $\b$--deformations of \cite{LS}. In
this construction, one starts from the $\cN=4$ theory with
complexified gauge coupling
\be
\label{2-1}
\tau = {\vartheta_{\rm YM} \ov 2\pi} + {4 \pi {\rm i}
\ov g_{\rm YM}^2}\ ,
\ee
and applies a deformation that acts on the three complex
chiral superfields $\Phi_i$, $i=1,2,3$, of the theory by modifying
the standard superpotential $W={\rm Tr} ( \Phi_1 [ \Phi_2 , \Phi_3
] )$ to
\be
\label{2-2}
W = {\rm Tr}(e^{{\rm i}\pi\b} \Phi_1 \Phi_2 \Phi_3 - e^{-{\rm i}\pi\b} \Phi_1 \Phi_3 \Phi_2)\ ,
\ee
where $\b$ is a complex phase. The latter is conveniently
parametrized as $\b = \g - \tau \s$, where $\g$ and $\s$ are real
parameters with unit period; in the special cases $\s=0$ or
$\g=0$, the deformation is referred to as a $\g$--deformation or a
$\s$--deformation respectively. The above deformation breaks
$\cN=4$ supersymmetry down to $\cN=1$ and the corresponding
$SO(6)_\cR$ global $\cR$--symmetry group down to its $U(1)_1
\times U(1)_2 \times U(1)_\cR$ Cartan subgroup where $U(1)_\cR$
stands for the surviving R-symmetry. There is also a
$\mathbb{Z}_3$ symmetry under cyclic permutations. Under
$U(1)_1 \times U(1)_2$, the charges of the chiral superfields
are
\be
\label{2-3}
(Q_1^{\Phi_1},Q_1^{\Phi_2},Q_1^{\Phi_3})=(0,1,-1)\ ,\qq (Q_2^{\Phi_1},Q_2^{\Phi_2},Q_2^{\Phi_3})=(-1,1,0)\ ,
\ee
while the superpotential is invariant. The Coulomb branch of the
theory is described by the F-term conditions
\ba
\label{2-4}
\Phi_1 \Phi_2 = q \Phi_2 \Phi_1\ ,\qq q\equiv e^{-2 {\rm i} \pi
\b}\ ,\qq  {\rm and\ cyclic}\ ,
\ea
which are valid for large $N$ (exact for $U(N)$). For generic
$\b$, these conditions are solved by traceless $N \times N$
matrices, where in each entry at most one of them is nonzero. For
$\g$--deformations with rational $\g$, the Coulomb branch contains
additional regions \cite{berenstein,dorey}.

\no On the gravity side, the dual to the Leigh-Strassler
deformation was constructed by Lunin and Maldacena in \cite{LM}
for field theories possessing at least a $U(1)_1 \times U(1)_2$
global symmetry which, in the gravity dual, corresponds to an
isometry of the supergravity background along two angular
directions parametrizing a 2-torus. The deformation proceeds by
applying a certain $\b$--dependent $SL(3,\mathbb{R})$
transformation belonging to the first factor of the full
$SL(3,\mathbb{R}) \times SL(2,\mathbb{R})$ duality group of Type
IIB supergravity on the 2-torus. The effect of this transformation
in the field theory amounts to the replacement of the standard
product of field operators by the Moyal-like product
\be
\label{2-4a}
f \ast g = e^{{\rm i} \pi \b ( Q_1^f Q_2^g - Q_1^g Q_2^f )} f g\ ,
\ee
which, for the case of $\cN=4$ SYM, does indeed induce the
modified superpotential \eqn{2-2}. The transformation outlined
above was applied to various Type IIB backgrounds in \cite{LM} and
further generalized to a broader class of backgrounds in
\cite{marginal-otherbackgrounds}.

\no To illustrate the construction for the solutions of interest,
we consider a general Type IIB background where the ten spacetime
coordinates are split into a seven-dimensional part parametrized
by $x^I$, $I=1,2,\dots,7$ and a three-dimensional part
corresponding to a 3-torus parametrized by the angles $\phi_i$,
$i=1,2,3$. The most general metric of this form is given by
\be
\label{2-5}
ds^2_{10} = G_{IJ}(x) dx^I dx^J
+ 2 \sum_{i=1}^3 \lambda_{Ii}(x) dx^I d \phi_i + \sum_{i=1}^3 z_i(x) d \phi_i^2 \ ,
\ee
where the $z_i$ are positive-definite functions.
This metric contains the mixed $dx^Id\phi_i$-terms and generalizes
that considered in \cite{hsz} as the staring point for constructing marginally deformed
backgrounds.
Apart from the
metric, the solution is characterized by the dilaton and the RR
4-form
\be
\label{2-8}
A_4 = C_4 + C_3^{i} \wedge d \phi_i + \frac{1}{2}
\epsilon_{ijk} C_2^{i} \wedge d \phi_j \wedge d \phi_k + C_1
\wedge d \phi_1 \wedge d \phi_2 \wedge d \phi_3\ ,
\ee
where $C_1$, $C_2^i$, $C_3^i$ and $C_4$ are forms of degree
indicated by the lower index. They  have dependence and support
only on $x^I$ and are constrained by the self-duality relations
\be
\label{2-9} d C_1 \wedge d \phi_1 \wedge d \phi_2 \wedge d \phi_3
= \star d C_4\ , \qq \frac{1}{2} \epsilon_{ijk} d C_2^i \wedge d
\phi_j \wedge d \phi_k = \star (d C_3^i \wedge d \phi_i)\ .
\ee
A special case of the above solution, which includes the
rotating branes to be examined in Section 3, is the one
where the only nonzero $\l_{Ii}$ are those in the $ti$ directions,
$\l_i \equiv \l_{ti}$. For this case, the metric \eqn{2-5}
simplifies to
\be
\label{2-10}
ds^2_{10} = G_{IJ}(x) dx^I dx^J + 2 \sum_{i=1}^3 \lambda_i(x) dt d \phi_i + \sum_{i=1}^3 z_i(x) d \phi_i^2 \ .
\ee
According to the gauge/gravity correspondence, the three chiral
superfields $\Phi_i$ are in one-to-one correspondence with three
complex coordinates $w_i = R_i(x) e^{{\rm i} \phi_i}$ and the
generators $(Q_1,Q_2,Q_\cR)$ of the global symmetries of the gauge
theory correspond to linear combinations of the generators
$(J_{\phi_1},J_{\phi_2},J_{\phi_3})$ of the shifts along the
torus. To proceed, it is most convenient to trade the $\phi_i$ for
a new set of variables $\vphi_i$ such that the generators
$(J_{\vphi_1},J_{\vphi_2},J_{\vphi_3})$ of shifts along these
directions become precisely identified with $(Q_1,Q_2,Q_\cR)$.
From \eqn{2-3}, it is easily seen that the appropriate set of
variables is
\be
\label{2-11}
\varphi_1={1\ov 3}(\phi_1+\phi_2-2\phi_3)\ , \qq
\varphi_2={1\ov 3}(\phi_2+\phi_3-2 \phi_1)\ , \qq \varphi_3={1\ov
3}(\phi_1+\phi_2+\phi_3)\ .
\ee
In terms of these new variables, the description of the
$SL(3,\mathbb{R})$ transformations corresponding to
$\b$--deformations in the gravity dual is rather simple. These
transformations can be written \cite{frolov} as the sequence ${\it
STsTS}^{-1}$ where ${\it T}$ stands for a T--duality along the
isometry direction $\vphi_1$, ${\it S}$ and ${\it S}^{-1}$ stand
for an S--duality with parameter $\tilde{\s} \equiv \s / \g$ and
$-\tilde{\s}$ respectively (see the appendix for conventions), and
${\it s}$ denotes the coordinate shift
\be
\label{2-12} {\textit s}\  : \qq \varphi_2 \to \varphi_2+\g
\varphi_1\ .
\ee
The special case of $\g$--deformations corresponds to the sequence
${\it TsT}$ and is obtained from the above case by taking
$\tilde{\s}=0$. The detailed procedure is presented below.

\subsection{Construction of the marginally-deformed solutions}

We now construct the $\b$--deformations of the gravity duals under
consideration by applying the ${\it STsTS}^{-1}$ sequence of
transformations to the solution \eqn{2-5}--\eqn{2-9}. Here, we will
denote the metric components in the $\phi$ and $\vphi$ bases by
$G_{MN}$ and $g_{MN}$ respectively, while we will indicate the
various fields at the intermediate steps by superscripts, e.g.
$g^{(1)}_{MN}, \, g^{(2)}_{MN}, \, \ldots \,$, and those at the
final step by a hat. In the various computations we use the T-- and
S--duality rules that we have written in the appendix in a compact
way particularly useful for our purposes. To present the results in
a succinct form, it is convenient to introduce the 2-forms
\ba
\label{2-13}
\cB_2 &=& \l_{I1} \l_{J2} dx^I \wedge dx^J + z_1 (\l_{I2} - \l_{I3})
d\phi_1 \wedge dx^I + z_1 z_2 d\phi_1 \wedge d\phi_2 + {\rm cyclic}\ , \nonumber\\
\cA_2 &=& C_1 \wedge (d\phi_1 + d\phi_2 + d\phi_3) + C_2^1 + C_2^2 + C_2^3\ ,
\ea
the first of which can be written as the product of two one-forms
$\cB_2 = \cA_1 \wedge \cB_1$ with
\ba
\label{2-14}
\cA_1 &=& (\l_{I2} - \l_{I3}) dx^I + z_2 d \phi_2 - z_3 d\phi_3\ , \nonumber\\
\cB_1 &=&  \left( {\l_{I2} z_3 + \l_{I3} z_2 \ov z_2+z_3} - \l_{I1} \right) dx^I - z_1 d \phi_1 + {z_2 z_3 \ov z_2+z_3} (d\phi_2+d\phi_3)\ ,
\ea
and hence is nilpotent in the sense that
\be
\label{2-15}
\cB_2 \wedge \cB_2 = 0 \ .
\ee
It is also useful to introduce
\ba
\label{2-16}
\cG^{-1} &=& 1 + | \b |^2  (z_1 z_2 + z_2 z_3 + z_3 z_1)\ , \nonumber\\
\cH &=& 1 + \s^2 e^{-2 \Phi} (z_1 z_2 + z_2 z_3 + z_3 z_1)\ ,\\
\cQ &=& \g \s e^{-\Phi}  (z_1 z_2 + z_2 z_3 + z_3 z_1)\ ,
\nonumber
\ea
where
\be
\label{2-17}
\b = \g - \tau \s = \g - {\rm i} \s e^{- \Phi}\ ,
\ee
since the axion in the initial solution is zero. The calculation
proceeds as follows:

\no $\bullet$ In the first step we perform an S--duality with
parameter $\tilde{\s} = \s / \g$. Applying the transformation and
passing to the $\vphi$ basis, we obtain the metric
\ba
\label{2-18}
&&g^{(1)}_{IJ} = {|\b| \ov \g} G_{IJ} , \nonumber\\
&&g^{(1)}_{I1} = {|\b| \ov \g} (\l_{I2} - \l_{I3})\ ,
\qq g^{(1)}_{I2} = {|\b| \ov \g}(\l_{I2} - \l_{I1})\ ,
\qq  g^{(1)}_{I3} = {|\b| \ov \g} (\l_{I1} + \l_{I2} + \l_{I3})\ ,
\nonumber\\
&&g^{(1)}_{11} = {|\b| \ov \g} (z_2+z_3)\ ,\qq g^{(1)}_{22} =
{|\b| \ov \g} (z_1 + z_2)\ ,\qq g^{(1)}_{33} = {|\b| \ov \g}(z_1 +
z_2 + z_3)\ ,
\\
&&g^{(1)}_{12}= {|\b| \ov \g} z_2\ ,\qq g^{(1)}_{13} = {|\b| \ov
\g} (z_2 - z_3)\ ,\qq g^{(1)}_{23}= {|\b| \ov \g} (z_2 - z_1)\ ,
\nonumber
\ea
the dilaton
\be
\label{2-19}
e^{2\Phi^{(1)}}= {|\b|^4 \ov \g^4} e^{2\Phi}\ ,
\ee
and the axion
\be
\label{2-20}
A_0^{(1)} = - {\g \s \ov |\b|^2} e^{-2\Phi}\ .
\ee
The RR 4-form remains unchanged and, in the new basis, reads
\ba
\label{2-21}
&&\!\!\!\!\!\!\!\!A^{(1)}_4 = \cA_2\wedge d \vphi_1 \wedge d \vphi_2
+ (C_2^1 + C_2^2 - 2 C_2^3) \wedge d \vphi_2 \wedge d \vphi_3 +
(C_2^2 + C_2^3 - 2 C_2^1) \wedge d \vphi_3 \wedge d \vphi_1
\nonumber\\ &&\!\!\!\!\!\!\!\! \qq + \:\: (C_3^2 - C_3^3) \wedge d \vphi_1 +
(C_3^2 - C_3^1) \wedge d \vphi_2 + (C_3^1 + C_3^2 + C_3^3) \wedge
d \vphi_3 + C_4\ ,
\ea
while the NSNS and RR 2-forms are zero.

\no
$\bullet$
In the second step we perform a
T--duality along $\vphi_1$. The NSNS fields are given by
\ba
\label{2-22}
&&g^{(2)}_{IJ} = {|\b| \ov \g} \left( G_{IJ} - {(\l_{I2} - \l_{I3})(\l_{J2} - \l_{J3}) \ov z_2 + z_3} \right)\ , \nonumber\\
&&g^{(2)}_{I2} = {|\b| \ov \g} \left( {z_2 \l_{I3} + z_3 \l_{I2}
\ov z_2 + z_3} - \l_{I1} \right)\ ,
\qq g^{(2)}_{I3} = {|\b| \ov \g} \left( {2 ( z_2 \l_{I3} + z_3 \l_{I2} ) \ov z_2 + z_3} + \l_{I1} \right)\ , \nonumber\\
&&g^{(2)}_{11} = {\g \ov |\b|} {1\ov z_2+z_3}\ , \qq g^{(2)}_{22}
= {|\b| \ov \g} {z_1 z_2 + z_2 z_3 + z_3 z_1 \ov z_2+z_3}\ ,\qq
g^{(2)}_{33}= {|\b| \ov \g} \left( z_1+{4z_2z_3\ov z_2+z_3}
\right)\ ,
\nonumber\\
&&g^{(2)}_{23}= {|\b| \ov \g} \left(  {2z_2z_3\ov z_2+z_3}-z_1 \right)\ ,
\\
&&B^{(2)}_{2} = - {1 \ov z_2+z_3} \left[ (\l_{I2}-\l_{I3}) dx^I + z_2 d\vphi_2 + (z_2-z_3) d\vphi_3 \right] \wedge d\vphi_1\ , \nonumber\\
&&e^{2\Phi^{(2)}}= {|\b|^3 \ov \g^3} {e^{2\Phi} \ov z_2+z_3}\ ,
\nonumber
\ea
while the RR fields are
\ba
\label{2-23}
&&A^{(2)}_1 = - {\g \s \ov |\b|^2} e^{-2\Phi} d\vphi_1\ , \nonumber\\
&&A^{(2)}_3 = - \cA_2 \wedge d\vphi_2 + (C_2^2+C_2^3-2C_2^1) \wedge d\vphi_3 + (C_3^2 - C_3^3)\ ,
\\
&&A^{(2)}_5 = A^{(1)}_4 \wedge d\vphi_1 + B^{(2)}_{2} \wedge A^{(2)}_3\ .
\nonumber
\ea

\no $\bullet$ In the third step we perform a coordinate shift
$\varphi_2 \to \varphi_2+\g \varphi_1$. The changed metric
components are
\ba
\label{2-24}
&&g^{(3)}_{I1}= | \b | \left( {z_2 \l_{I3} + z_3
\l_{I2} \ov z_2 + z_3} - \l_{I1} \right)\ ,
\nonumber\\
&&g^{(3)}_{11}= {\g \ov |\b|} \cG^{-1} {1 \ov z_2+z_3}\ ,
\\
&&g^{(3)}_{12} = |\b| {z_1 z_2 + z_2 z_3 + z_3 z_1 \ov z_2+z_3}\ ,\qq
g^{(3)}_{13}= |\b| \left({2 z_2 z_3\ov z_2+z_3}-z_1\right)\ .
\nonumber
\ea
The 3-form changes as
\be
\label{2-25}
A^{(3)}_3 = A^{(2)}_3 - \g \cA_2 \wedge d \vphi_1\ ,
\ee
while all other fields remain invariant.

\no $\bullet$ In the fourth step another T--duality along
$\vphi_1$ is performed. Applying the duality and returning to the
$\phi$ basis, we find that the NSNS fields are given by
\ba
\label{2-26}
&&G^{(4)}_{IJ} = {|\b| \ov \g} \left\{ G_{IJ} - |\b|^2 \cG
\left[ z_1 (\l_{I2} - \l_{I3})(\l_{J2} - \l_{J3}) + {\rm cyclic} \right] \right\}\ , \nonumber\\
&&G^{(4)}_{Ii} = {|\b| \ov \g} \cG \left[ \l_{Ii} + |\b|^2 ( z_1 z_2 \l_{I3} + {\rm cyclic} ) \right]\ ,
\nonumber\\
&&G^{(4)}_{ij} = {|\b| \ov \g} \cG ( z_i \d_{ij} + |\b|^2 z_1 z_2 z_3 )\ ,
\\
&&B^{(4)}_2 = {|\b|^2 \ov \g} \cG \cB_2\ ,
\nonumber\\
&&e^{2\Phi^{(4)}}= {|\b|^4 \ov \g^4} \cG e^{2\Phi}\ ,
\nonumber
\ea
while the RR fields are
\ba
\label{2-27}
&&A_0^{(4)} = - {\g \s \ov |\b|^2} e^{-2\Phi}\ , \qq A^{(4)}_2 = - \g \cA_2 - \s e^{-2 \Phi} \cG \cB_2\ ,
\nonumber\\
&&A^{(4)}_4 = A_4 - |\b|^2 \cG \cB_2 \wedge \cA_2\ , \qq A^{(4)}_6 = {|\b|^2 \ov \g} \cG A_4 \wedge \cB_2\ .
\ea

\no
$\bullet$
The final step is another S--duality, now with parameter $-\tilde
\s$. This leads to the gravity dual of the $\b$--deformed theory,
expressed in terms of the NSNS fields
\ba
\label{2-28}
\hat{G}_{IJ} &=& \cH^{1/2} \left\{ G_{IJ} - |\b|^2 \cG
\left[ z_1 (\l_{I2} - \l_{I3})(\l_{J2} - \l_{J3}) + {\rm cyclic} \right] \right\}\ , \nonumber\\
\hat{G}_{Ii} &=& \cG \cH^{1/2}  \left[ \l_{Ii} + |\b|^2 ( z_1 z_2 \l_{I3} + {\rm cyclic} ) \right]\ ,
\nonumber\\
\hat{G}_{ij} &=& \cG \cH^{1/2} ( z_i \d_{ij} + |\b|^2 z_1 z_2 z_3 )\ ,
\\
\hat{B}_2 &=& \g \cG \cB_2 - \s \cA_2\ , \nonumber\\
e^{2\hat{\Phi}} &=& \cG \cH^2 e^{2\Phi}\ ,
\nonumber
\ea
and the RR fields
\ba
\label{2-29}
\hat{A}_0 &=& \cH^{-1} \cQ e^{-\Phi}\ , \nonumber\\
\hat{A}_2 &=&  - \g \cA_2 - \s e^{-2 \Phi} \cG \cB_2\ , \\
\hat{A}_4 &=& A_4 - \g^2 \cG \cB_2 \wedge \cA_2 + {1 \ov 2} \g \s \cA_2 \wedge \cA_2 \nonumber\ ,
\\
\hat{A}_6
&=& \hat{B}_2 \wedge \hat{A}_4\ ,
\nonumber
\ea
where in the last relation we made use of the nilpotency of
$\cB_2$ \eqn{2-15}. One can check that the above formulas reduce,
for the appropriate limiting cases, to the various solutions that
have been found in the literature \cite{LM,frolov,hsz}. The case
of $\g$--deformations is obtained as the special case of the above
where $\s=0$, in which case the quantities in Eq. \eqn{2-16}
reduce to
\be
\label{2-30}
\cG^{-1} = 1 + \g^2 (z_1 z_2 + z_2 z_3 + z_3 z_1)\ ,\qq \cH = 1\ ,\qq \cQ = 0\ .
\ee

\no At this point, it is instructive to compare the general case
of $\b$--deformations with the special case of $\g$--deformations.
We see that the extra $\s$--dependence in the former enters
through (i) the replacement $\g^2 \to |\b|^2$ in the metric and in
the definition of $\cG$, (ii) the overall factors $\cH^{1/2}$ and
$\cH^2$ in the metric and dilaton respectively and (iii) a nonzero
axion as well as new terms in the NSNS 2-form, the RR 2-form and
the RR 4-form proportional to $\cA_2$, $\cB_2$ and $\cA_2 \wedge
\cA_2$, respectively. These changes clearly affect the Nambu--Goto
(or Dirac--Born--Infeld plus Wess--Zumino) actions of probe
strings (or branes) propagating in the deformed geometry and so,
in relevant investigations, one is entitled to expect qualitative
departures from results obtained for purely $\g$--deformed
backgrounds. On the other hand, one readily verifies that the
massless scalar wave equation $\del_M ( \sqrt{-G} e^{-2 \Phi}
G^{MN} \del_N \Psi) = 0$ is insensitive to the presence of the
$\cH$ factors, which implies that its analysis proceeds as in the
$\g$--deformed case with the replacement $\g^2 \to |\b|^2$ and
that, in the case of multicenter D3-branes considered in
\cite{hsz}, its relation with integrable systems found there
remains intact.

\section{Deformations of rotating and multicenter D3-branes}
\label{sec3}

In this section, we explicitly apply the $\b$--deformation
procedure described above to rotating D3-brane solutions. First,
we give a brief review of the field-theory limit of these
solutions and we focus on three special cases where the metrics
simplify considerably. Then, we present the full metrics of the
corresponding deformed solutions. Finally, we demonstrate that the
thermodynamic properties of the deformed metrics are exactly the
same as for the undeformed ones.

\subsection{Rotating and multicenter D3-brane solutions}

The solutions we are interested in here are the non-extremal
rotating D3-branes found in full generality in \cite{trivedi}
using previous results from \cite{cy}. They are characterized by
the non-extremality parameter $\m$ plus the rotation parameters
$a_i$, $i=1,2,3$, which correspond to the three generators in the
Cartan subalgebra of $SO(6)$. The spacetime coordinates are split
into the brane coordinates $(t,\vec{x}_3)=(t,x_1,x_2,x_3)$ and the
transverse coordinates $y_m$, $m=1,\ldots,6$, which can be
parametrized as
\ba
\label{3-1}
w_1 &=& y_1 + {\rm i} y_2 = \sqrt{r^2+a_1^2} \sin \th e^{{\rm i} \phi_1}\ , \nonumber\\
w_2 &=& y_3 + {\rm i} y_4 = \sqrt{r^2+a_2^2} \cos \th \sin \psi e^{{\rm i} \phi_2}\ , \nonumber\\
w_3 &=& y_5 + {\rm i} y_6 = \sqrt{r^2+a_3^2} \cos \th \cos \psi e^{{\rm i} \phi_3}\ ,
\ea
where the complex coordinates $w_i$ are in one-to-one
correspondence with the chiral superfields $\Phi_i$ of the gauge
theory. Here we are interested in the field-theory limit of these
solutions which, in the most general non-extremal case, is given
by the metric (we follow \cite{rs}, in which the thermodynamic properties of the
solution, presented below in some special cases, were also computed)
\ba
\label{3-2}
&& ds^2 = H^{-1/2} \left[-\left(1-{\m^4\ov r^4 \D}\right) dt^2 + d \vec{x}_3^2 \right]
+ H^{1/2}{r^6 \D \ov f}\ dr^2
\nonumber\\
&& \, + \:\: H^{1/2}
\Bigg[r^2 \Delta_1 d\theta^2 + r^2 \Delta_2 \cos^2\theta d\psi^2 +
2 (a_2^2-a_3^2)\cos\theta\sin\theta\cos\psi\sin\psi d\theta d\psi
\\
&& \, + \:\: (r^2+a_1^2) \sin^2\theta d\phi_1^2 +
(r^2+a_2^2) \cos^2\theta \sin^2\psi d\phi_2^2 +
(r^2+a_3^2) \cos^2\theta\cos^2\psi d\phi_3^2
\nonumber\\
&& -\ 2 {\m^2\ov R^2} \ dt \ (a_1 \sin^2 \th\ d \phi_1 + a_2 \cos^2 \th \sin^2\psi d \phi_2
+ a_3 \cos^2 \th \cos^2\psi d \phi_3 )
\Bigg] \ ,
\nonumber
\ea
the constant dilaton
\be
\label{2-6} e^\Phi = e^{\Phi_0} = g_{\rm s}\ ,
\ee
and a 4-form potential of the form \eqn{2-9} with
\ba
\label{3-5} \!\!\!\!\!\!\!\!\!\!\!\!C_4 &=& - {H^{-1} \ov g_{\rm
s}} dt \wedge dx_1 \wedge dx_2 \wedge dx_3\ ,
\nonumber\\
\!\!\!\!\!\!\!\!\!\!\!\!(C_3^1,\, C_3^2,\, C_3^3) &=& - {\m^2 \ov
R^2 g_{\rm s}} (a_1 \sin^2 \th,\, a_2 \cos^2 \th \sin^2 \psi,\,
a_3 \cos^2 \th \cos^2 \psi) dx_1 \wedge dx_2 \wedge dx_3\ ,
\ea
and $C_1$ and $C_2^i$ specified by the duality relations
\eqn{2-9}. In the above, the various functions are given by
\ba
\label{3-3}
H & = & {R^4\ov r^4 \D}\ ,
\nonumber\\
f & = & (r^2+a_1^2)(r^2+a_2^2)(r^2+a_3^2)- \m^4 r^2\ ,
\nonumber\\
\Delta &=& 1 +{a_1^2\over r^2} \cos^2\theta +{a_2^2\over r^2}
(\sin^2\theta\sin^2\psi +\cos^2\psi )
+ {a_3^2\over r^2}(\sin^2\theta\cos^2\psi +\sin^2\psi )
\nonumber \\
&+& {a_2^2 a_3^2\over r^4}\sin^2\theta +{a_1^2 a_3^2\over r^4}
\cos^2\theta\sin^2\psi +{a_1^2 a_2^2 \over r^4}\cos^2\theta\cos^2\psi\ ,
\\
\Delta_1 &=& 1+{a_1^2\ov r^2}\cos^2\theta +
{a_2^2\ov r^2}\sin^2\theta\sin^2\psi +
{a_3^2\ov r^2}\sin^2\theta\cos^2\psi\ ,
\nonumber\\
\Delta_2 &=& 1+{a_2^2\ov r^2}\cos^2\psi +{a_3^2\ov r^2}\sin^2\psi\ .
\nonumber
\ea
In the notation of Section 2, the metric components $z_i$ and
$\l_i$ are given by
\ba
\label{3-4}
\!\!\!\!(z_1,\, z_2,\, z_3) &=& {R^2 \ov r^2 \D^{1/2}} \left( (r^2+a_1^2) \sin^2 \th,\,
(r^2+a_2^2) \cos^2 \th \sin^2 \psi,\,(r^2+a_3^2) \cos^2 \th \cos^2 \psi \right)\ ,\nonumber\\
\!\!\!\!(\l_1,\, \l_2,\, \l_3) &=& - { \m^2 \ov r^2 \D^{1/2}} ( a_1 \sin^2
\th,\, a_2 \cos^2 \th \sin^2 \psi,\, a_3 \cos^2 \th \cos^2 \psi)\ ,
\ea
For $\m \ne 0$, these solutions describe rotating branes and are
dual to $\cN=4$ SYM at finite temperature and R-charge chemical
potentials, and the $a_i$ parametrize the angular
velocities/momenta on the supergravity side and the chemical
potentials/R-charges on the gauge theory-side. For $\m = 0$, these
solutions describe multicenter brane distributions dual to the
Coulomb branch of $\cN=4$ SYM, and the $a_i$ parametrize the
principal radii of the distribution on the supergravity side and
the scalar {\em vevs} on the gauge-theory side. In the rest of the
paper, we will refer to the $a_i$ as ``rotation parameters'',
keeping in mind their different interpretations in these two
cases.

\no In what follows, we examine some simple special cases of the
above general solution, namely those corresponding to three equal
nonzero rotation parameters, two equal nonzero rotation parameters
and one nonzero rotation parameter.

\subsubsection*{Three equal rotation parameters}

The first special case we consider is the one where all three
rotation parameters are equal to each other, $a_1=a_2=a_3=r_0$.
Employing the change of variable $r^2 \to r^2 - r_0^2$ we write the
resulting metric as
\ba
\label{3-6}
&&ds^2 = H^{-1/2} \left[ - \left( 1 - {\m^4 \ov r^4} \right) dt^2 + d \vec{x}_3^2 \right]
+ H^{1/2}{r^6 \ov r^6 - \m^4 (r^2-r_0^2)}\ dr^2
\nonumber\\
&&\, + \:\: H^{1/2}
\left\{ r^2 d \Omega_5^2 - {2 \m^2 r_0 \ov R^2} dt \left[ \sin^2 \th d \phi_1
+ \cos^2 \th ( \sin^2 \psi d \phi_2 + \cos^2 \psi d \phi_3) \right] \right\}\ ,
\ea
where
\be
\label{3-7}
H = {R^4 \ov r^4}\ ,
\ee
and $d \Omega_5^2$ is the ${\rm S}^5$ metric
\be
\label{3-8}
d \Omega_5^2 = d\th^2 + \sin^2 \th d\phi_1^2 + \cos^2 \th ( d\psi^2 + \sin^2 \psi d\phi_2^2 + \cos^2 \psi d\phi_3^2) \ .
\ee
The horizon radius $r_H$ for this metric is given by the largest
root of the equation
\be
\label{e-3r}
r^6-\m^4(r^2-r_0^2)=0\ ,
\ee
while its Hawking temperature reads
\be
\label{e-3t}
T = {r_H(2 r_H^2 - 3 r_0^2) \ov 2 \pi R^2(r_H^2 - r_0^2)}\ .
\ee
For $\m=0$ this background reduces to the ${\rm AdS}_5 \times {\rm
S}^5$ background obtained by stacked D3-branes at the
origin.

\subsubsection*{Two equal rotation parameters}

The second special case is the one where two rotation parameters
are set to the same nonzero value, which we may take as $a_2 = a_3
= r_0$. Employing again the change of variable $r^2 \to r^2-r_0^2$
we have the metric
\ba
\label{3-9}
&& ds^2 = H^{-1/2} \left[-\left(1-{\m^4 H \ov R^4}\right) dt^2 + d \vec{x}_3^2 \right]
+ H^{1/2} {r^4(r^2-r_0^2 \cos^2\th)\ov (r^4-\m^4)(r^2-r_0^2)}\ dr^2
\nonumber\\
&& + H^{1/2}\Big[(r^2-r_0^2\cos^2\th )d\th^2 + r^2 \cos^2\th d\Om_3^2
+ (r^2-r_0^2)\sin^2\th d \phi_1^2
\\
&& - \ 2 {\m^2 r_0\ov R^2} \ dt \cos^2\th (\sin^2\psi d \phi_2 + \cos^2\psi d \phi_3)\Big]\ ,
\nonumber
\ea
where
\be
\label{3-10}
H = {R^4 \ov r^2 (r^2 - r_0^2 \cos^2 \th )}\ ,
\ee
while $d \Omega_3^2$ is the ${\rm S}^3$ metric
\be
\label{3-11}
d \Omega_3^2 = d\psi^2 + \sin^2 \psi d\phi_2^2 + \cos^2 \psi d\phi_3^2\ .
\ee
Now, the horizon radius is simply
\be
r_H = \m\ ,
\ee
and the Hawking temperature reads
\be
T = {\sqrt{\m^2 - r_0^2} \ov \pi R^2}\ .
\ee
For $\m=0$ this background reduces to that obtained by a uniform
distribution of D3-branes on a 3-sphere of radius $r_0$.

\subsubsection*{One rotation parameter}

The third special case is the one where there is only one nonzero
rotation parameter, which we may take as $a_1 = r_0$. In this
case, we have the metric
\ba
\label{3-12}
&& ds^2 = H^{-1/2} \left[-\left(1-{\m^4 H \ov R^4}\right) dt^2 + d \vec{x}_3^2 \right]
+ H^{1/2} {r^2(r^2+r_0^2 \cos^2\th)\ov r^4+r_0^2 r^2 -\m^4}\ dr^2
\nonumber\\
&& + H^{1/2}\Big[(r^2+r_0^2\cos^2\th )d\th^2 + r^2 \cos^2\th d\Om_3^2
+ (r^2+r_0^2)\sin^2\th d \phi_1^2
\\
&& - \ 2 {\m^2 r_0\ov R^2}  \sin^2\th dt d\phi_1 \Big]\ ,
\nonumber
\ea
where the harmonic function is given by
\be
\label{3-13}
H = {R^4 \ov r^2(r^2 + r_0^2 \cos^2 \th)}\ ,
\ee
while $d \Omega_3^2$ is defined as before. The horizon radius is given by
\be
r_H^2 = {1 \ov 2} \left( - r_0^2 + \sqrt{r_0^4 + 4 \m^4} \right)\ ,
\ee
and the Hawking temperature reads
\be
T = {r_H \sqrt{r_0^4 + 4 \m^4} \ov 2 \pi R^2 \m^2}\ .
\ee
For $\m=0$ this background reduces to that obtained by a uniform
distribution of D3-branes on a disc of radius $r_0$ and is related
to the corresponding background for two rotation parameters by the
transformation $r_0^2 \to -r_0^2$.

\subsection{The deformed metrics}

After the above preliminaries, we are ready to apply the
marginal-deformation procedure to the rotating-brane solutions just
discussed. In what follows, we present the explicit form of the
marginally-deformed metrics for the three special cases considered
earlier. To keep the notation as compact as possible, it is
convenient to introduce the shorthand notation $\calpha \equiv \cos
\alpha$ and $\salpha \equiv \sin \alpha$ and the following rescaling
for the deformation parameters\footnote{These parameters differ by
the analogous parameters in \cite{hsz} by a factor of $1/2$.}
\be
\label{3-14} \hbeta \equiv {R^2 \b \ov 2} \ , \qq \hgamma \equiv
{R^2 \g \ov 2}\ ,\qq \hsigma \equiv {R^2 \s \ov 2 g_{\rm s}}\ ,
\ee
with the new parameters satisfying $|\hbeta|^2 = \hgamma^2 +
\hsigma^2$.

\subsubsection*{The zero-rotation case}

As a first example, let us consider the case where all rotation
parameters are set to zero. Setting $r_0=0$ in any of the above
three metrics and substituting in \eqn{2-27}, we find
\be
\label{3-16}
ds^2 = \cH^{1/2} \left\{ H^{-1/2} \left[ - \left( 1 - {\m^4 \ov r^4} \right) dt^2 + d \vec{x}_3 ^2 \right] + H^{1/2} \left( {r^4 \ov r^4 - \m^4}  dr^2 + r^2 d\Omega_{5,\b}^2 \right) \right\}\ ,
\ee
where $d\Omega_{5,\b}^2$ is the metric on the deformed five-sphere
${\rm S}^5_\b$, given by
\be
\label{3-17}
d\Omega_{5,\b}^2 = d \th^2 + \cG \sth^2 d \phi_1^2 + \cth^2 [ d \psi^2  + \cG ( \spsi^2 d\phi_2^2 + \cpsi^2 d \phi_3^2 ) ] + \cG |\hbeta|^2 \cth^4 \sth^2 {\rm s}_{2 \psi}^2 \left(\sum_{i=1}^3 d\phi_i \right)^2\ ,
\ee
and the various functions are
\be
\label{3-18}
\cG^{-1} = 1 + 4 |\hbeta|^2 \cth^2 (\sth^2 + \cth^2 \cpsi^2 \spsi^2)\ , \qq
\cH = 1 + 4\hsigma^2 \cth^2 (\sth^2 + \cth^2 \cpsi^2 \spsi^2)\ ,
\ee
and $H=R^4/r^4$. The resulting space is thus a conformal rescaling
of the product of ${\rm AdS}_5$--Schwarzschild with ${\rm
S}^5_\b$.

\subsubsection*{Three equal rotation parameters}

Starting from the case with three equal rotation parameters, the
deformed metric is found by substituting the undeformed metric
\eqn{3-6} into Eq. \eqn{2-27}. After some algebra, we find
\ba
\label{3-21}
&&\!\!\!\!\!\!\!\!\!\!\!\!\!\!ds^2 = \cH^{1/2} H^{-1/2} \left\{ - \left[ 1 - {\m^4 [ r^2 - |\hbeta|^2 \cG r_0^2 \cth^2 ( \cth^4 {\rm s}_{2\psi}^2 + 4 \sth^4 + 4 \cth^2 \sth^2 {\rm c}_{4\psi}) ]  \ov r^6} \right] dt^2 + d \vec{x}_3 ^2 \right\}\nonumber\\
&&\!\!\!\!\!\!\!\!\!\!\!\!\!\!\, + \:\: \cH^{1/2} H^{1/2} \left[ {r^6 \ov r^6 - \m^4(r^2-r_0^2)}  dr^2 + r^2 d\Omega_{5,\b}^2 \right] \\
&&\!\!\!\!\!\!\!\!\!\!\!\!\!\!\, - \:\: \cG \cH^{1/2} H^{1/2} {2 \m^2 r_0 \ov R^2} dt \left[ \sth^2 d \phi_1 + \cth^2 ( \spsi^2 d\phi_2 + \cpsi^2 d \phi_3 ) + 3 |\hbeta|^2 \cth^4 \sth^2 {\rm s}_{2 \psi}^2 \sum_{i=1}^3 d\phi_i \right]\ , \nonumber
\ea
where $d\Omega_{5,\b}^2$ is as in \eqn{3-17}, $\cG^{-1}$ and $\cH$
are as in \eqn{3-18} and $H$ is as in \eqn{3-7}.

\subsubsection*{Two equal rotation parameters}

For the case with two equal rotation parameters, the deformed
metric is found to be
\ba
\label{3-23}
&&\!\!\!\!\!\!\!\!\!\!\!\!\!\!ds^2 = \cH^{1/2} H^{-1/2} \Biggl\{ - \biggl[ 1 - {\m^4 \Bigl[ 1 - {|\hbeta|^2 \cG r_0^2 \cth^4 \big( 4 (r^2-r_0^2) \sth^2 {\rm c}_{2\psi}^2 + r^2 \cth^2 {\rm s}_{2\psi}^2 \big) \ov r^2 (r^2 - r_0^2 \cth^2)} \Bigr]  \ov r^2 (r^2 - r_0^2 \cth^2)} \biggr] dt^2 + d \vec{x}_3 ^2 \Biggr\} \nonumber\\
&&\!\!\!\!\!\!\!\!\!\!\!\!\!\!\, + \:\: \cH^{1/2} H^{1/2} \left[ {r^4 (r^2 - r_0^2 \cth^2) \ov (r^2 - r_0^2) (r^4 - \m^4)}  dr^2 + (r^2 - r_0^2 \cth^2) d \th^2 + r^2 \cth^2 d \psi^2 \right] \\
&&\!\!\!\!\!\!\!\!\!\!\!\!\!\!\, + \:\: \cG \cH^{1/2} H^{1/2} \Biggl\{ (r^2 - r_0^2) \sth^2 d \phi_1^2 + r^2 \cth^2 ( \spsi^2 d\phi_2^2 + \cpsi^2 d \phi_3^2 ) + {|\hbeta|^2 r^2 (r^2-r_0^2) \cth^4 \sth^2 {\rm s}_{2 \psi}^2 \ov r^2 - r_0^2 \cth^2} \left(\sum_{i=1}^3 d\phi_i \right)^2 \nonumber\\
&&\!\!\!\!\!\!\!\!\!\!\!\!\!\!\, - \:\: {2 \m^2 r_0 \ov R^2} dt \left[ \cth^2 ( \spsi^2 d\phi_2 + \cpsi^2 d \phi_3 ) + {2 |\hbeta|^2 (r^2-r_0^2) \cth^4 \sth^2 {\rm s}_{2 \psi}^2 \ov r^2 - r_0^2 \cth^2} \sum_{i=1}^3 d\phi_i \right] \Biggr\}\ , \nonumber
\ea
with
\ba
\label{3-24}
\cG^{-1} &= & 1 +
4|\hbeta|^2 \cth^2\ { ( r^2 - r_0^2 ) \sth^2 + r^2 \cth^2 \cpsi^2 \spsi^2  \ov r^2 - r_0^2 \cth^2}\ ,
\nonumber\\
\cH & = & 1 + 4\hsigma^2 \cth^2\ { ( r^2 - r_0^2 ) \sth^2 + r^2 \cth^2 \cpsi^2 \spsi^2
 \ov r^2 - r_0^2 \cth^2}\ ,
\ea
and $H$ as in \eqn{3-10}.

\subsubsection*{One rotation parameter}

For the case with one rotation parameter, the deformed metric is
found to be
\ba
\label{3-27}
&&\!\!\!\!\!\!\!\!\!\!\!\!\!\!ds^2 = \cH^{1/2} H^{-1/2} \left\{ - \left[ 1 - {\m^4 ( r^2+r_0^2 \cth^2 - 4 |\hbeta|^2 \cG r_0^2 \cth^2 \sth^4 ) \ov r^2 (r^2 + r_0^2 \cth^2)^2} \right] dt^2 + d \vec{x}_3 ^2 \right\}\nonumber\\
&&\!\!\!\!\!\!\!\!\!\!\!\!\!\!\, + \:\: \cH^{1/2} H^{1/2} \left[ {r^2 (r^2 + r_0^2 \cth^2) \ov r^4 + r_0^2 r^2 - \m^4}  dr^2 + (r^2 + r_0^2 \cth^2) d \th^2 + r^2 \cth^2 d \psi^2 \right] \\
&&\!\!\!\!\!\!\!\!\!\!\!\!\!\!\, + \:\: \cG \cH^{1/2} H^{1/2} \Biggl[ (r^2 + r_0^2) \sth^2 d \phi_1^2 + r^2 \cth^2 ( \spsi^2 d\phi_2^2 + \cpsi^2 d \phi_3^2 ) + {|\hbeta|^2 r^2 (r^2+r_0^2) \cth^4 \sth^2 {\rm s}_{2 \psi}^2 \ov r^2 +   r_0^2 \cth^2} \left(\sum_{i=1}^3 d\phi_i \right)^2 \nonumber\\
&&\!\!\!\!\!\!\!\!\!\!\!\!\!\!\, - \:\: {2 \m^2 r_0 \ov R^2} dt \left( \sth^2 d\phi_1 + {|\hbeta|^2 r^2 \cth^4 \sth^2 {\rm s}_{2 \psi}^2 \ov r^2 + r_0^2 \cth^2} \sum_{i=1}^3 d\phi_i \right) \Biggr]\ , \nonumber
\ea
with
\ba
\label{3-28}
\cG^{-1} & = &  1 + 4|\hbeta|^2 \cth^2\ {( r^2 + r_0^2 )
\sth^2 + r^2 \cth^2 \cpsi^2 \spsi^2  \ov r^2 + r_0^2 \cth^2}\ ,
\nonumber \\
\cH & = & 1 + 4\hsigma^2 \cth^2\ { ( r^2 + r_0^2 ) \sth^2 + r^2
\cth^2 \cpsi^2 \spsi^2  \ov r^2 + r_0^2 \cth^2}\ ,
\ea
and $H$ as in \eqn{3-13}.

\subsection{Invariance of the thermodynamics}

A useful consistency check for our calculation is to examine the
thermodynamic properties of the deformed rotating-brane solutions.
In particular, since the deformed solutions are related to the
undeformed ones by U-duality transformations that are symmetries
of the underlying theory, all thermodynamic quantities for the
deformed metrics must be equal to those for the undeformed ones.
It is instructive to show that this is indeed the case by
explicitly calculating the angular velocities, the Hawking
temperature, and the entropy for the general deformed solutions.

\no We start by writing the deformation \eqn{2-28}, restricted to
a metric of the form \eqn{2-10}, in the Einstein frame. In this
frame, the deformed metric is $\hat{G}^{\rm (E)}_{MN} =
e^{-\hat{\Phi}/2} \hat{G}_{MN} = g_{\rm s}^{-1/2} \cG^{-1/4}
\cH^{-1/2} \hat{G}_{MN}$, and thus we have
\ba
\label{3-32}
\hat{G}^{\rm (E)}_{IJ} &=& g_{\rm s}^{-1/2} \cG^{-1/4} \{ G_{IJ} - |\b|^2 \cG \left[ z_1 (\l_{2} - \l_{3})^2 + {\rm cyclic} \right] \delta_{I,t} \delta_{J,t} \}\ , \nonumber\\
\hat{G}^{\rm (E)}_{ti} &=& g_{\rm s}^{-1/2} \cG^{3/4} [ \l_{i} + |\b|^2 ( z_1 z_2 \l_{3} + {\rm cyclic}) ]\ , \\
\hat{G}^{\rm (E)}_{ij} &=& g_{\rm s}^{-1/2} \cG^{3/4} (z_i \d_{ij}
+ |\b|^2 z_1 z_2 z_3 )\ , \nonumber
\ea
where we recall that, for the general rotating-brane solution
\eqn{3-2}, $z_i$ and $\l_i$ are given in \eqn{3-4} and $\cG$ and
$\cH$ are given in terms of the $z_i$ in \eqn{2-16}. Letting
$G^{\rm (E)}_{MN}=e^{-\Phi_0/2} G_{MN} = g_{\rm s}^{-1/2} G_{MN}$
be the undeformed metric in the Einstein frame, we write
\be
\label{3-31} \hat{G}^{\rm (E)}_{MN} = G^{\rm (E)}_{MN} + \d G^{\rm
(E)}_{MN}\ ,
\ee
where the functions $\d G^{\rm (E)}_{MN}$ represent the effect of
the deformation and are read off from \eqn{3-32} and the explicit
relations \eqn{3-4} for $z_i$ and $\l_i$ and \eqn{2-16} for $\cG$
and $\cH$. Although the resulting $\d G^{\rm (E)}_{MN}$ are very
complicated functions of $r$, $\th$ and $\psi$, inspection of
\eqn{2-16} and \eqn{3-4} shows that they satisfy
\be
\label{3-33} \d G^{\rm (E)}_{MN}|_{(\th,\psi)=(\pi/2,0)} = \d
G^{\rm (E)}_{MN}|_{(\th,\psi)=(0,\pi/2)} = \d G^{\rm
(E)}_{MN}|_{(\th,\psi)=(0,0)} = 0\ ,
\ee
while it can be shown that
\be
\label{3-34}
\partial_{r,\th,\psi} \d G^{\rm (E)}_{MN}|_{(\th,\psi)=(\pi/2,0)} = \partial_{r,\th,\psi} \d G^{\rm (E)}_{MN}|_{(\th,\psi)=(0,\pi/2)} = \partial_{r,\th,\psi} \d G^{\rm (E)}_{MN}|_{(\th,\psi)=(0,0)} = 0\ .
\ee
That is, there exist three values of $(\th,\psi)$ for which the
metric and its derivatives reduce to those in the undeformed case.

\no Given Eqs. \eqn{3-33} and \eqn{3-34}, it is very easy to check
that the angular velocities and the Hawking temperature are the
same as in the undeformed solution. Indeed, it immediately follows
that the horizon radius for the deformed metric is equal to the
horizon radius $r_H$ for the undeformed one. The angular
velocities $\hat{\Omega}_i$, are found by demanding that the
Killing vector $\xi = \partial_t + \hat{\Omega}_i
\partial_{\phi_i}$ associated with a stationary observer be null
at $r=r_H$, i.e. by solving the equation
\be
\label{3-35} \hat{\xi}^2 (r_H) = \xi^2(r_H) + \d G^{\rm
(E)}_{tt}(r_H) + 2 \d G^{\rm (E)}_{ti}(r_H) \hat{\Omega}_i + \d
G^{\rm (E)}_{ij}(r_H) \hat{\Omega}_i \hat{\Omega}_j = 0\ ,
\ee
where $\xi^2$ and $\hat{\xi}^2$ are the norms of $\xi$ with the
metrics $G^{\rm (E)}_{MN}$ and $\hat{G}^{\rm (E)}_{MN}$,
respectively. Evaluating this equation at $(\th,\psi)=(\pi/2,0)$,
$(0,\pi/2)$ and $(0,0)$ and using \eqn{3-33}, we obtain three
decoupled equations for $\hat{\Omega}_1$, $\hat{\Omega}_2$ and
$\hat{\Omega}_3$, respectively, which are the same ones that arise
for the undeformed metric \cite{rs}. Therefore, the angular
velocities are the same, $\hat{\Omega}_i = \Omega_i$. The Hawking
temperature is found from the relation
\be
\label{3-36} \hat{T}_H^2 = {1 \ov 16 \pi^2} \lim_{r \to r_H}
{\hat{G}^{{\rm (E)}MN} \partial_M \hat{\xi}^2 \partial_N
\hat{\xi}^2 \ov - \hat{\xi}^2}\ ,
\ee
which, being independent of the angles, can be evaluated at any of
the aforementioned three values of $(\th,\psi)$. Then, use of
\eqn{3-33} and \eqn{3-34} leads to the same relation as for the
undeformed metric and so $\hat{T}_H = T_H$.

\no Finally, the entropy is determined by the horizon area which
is in turn related to the determinant of the eight-dimensional
metric along the directions normal to the horizon. Labelling these
directions as $A=(\a,i)$ with $\a=(\vec{x}_3,\th,\psi)$ and
$i=(\phi_1,\phi_2,\phi_3)$, we find that this eight-dimensional
metric equals
\be
G^{\rm (E)}_{AB} = g_{\rm s}^{-1/2} \left( \begin{array}{cc}
G_{\alpha\beta} & 0
\\ 0 & z_i \d_{ij}
\end{array} \right)
\ee
and
\be
\hat{G}^{\rm (E)}_{AB} = g_{\rm s}^{-1/2} \left(
\begin{array}{cc} \cG^{-1/4} G_{\alpha\beta} & 0 \\ 0 & \cG^{3/4} (z_i
\d_{ij}+|\b|^2 z_1 z_2 z_3)
\end{array} \right)\ ,
\ee
for the undeformed and deformed cases, respectively. Then, a simple
calculation gives
\ba
\det \hat{G}^{\rm (E)}_{AB} &=& g_{\rm s}^{-4} \cG [1 + | \b |^2
(z_1 z_2 + z_2 z_3 + z_3 z_1)] z_1 z_2 z_3 \det G_{\alpha\beta}
\nonumber\\
&=& g_{\rm s}^{-4} z_1 z_2 z_3 \det G_{\alpha\beta} = \det G^{\rm
(E)}_{AB}\ ,
\ea
where in the second line we used the defining relation for $\cG$.
Therefore the entropy is indeed invariant, a fact that seems to be
closely related to the invariance of the central charge of the
dual CFT under deformations \cite{freedman}. This completes our
consistency check.

\section{Penrose limits of the $\b$--deformed solutions}
\label{sec4}

Having constructed the deformed solutions, it is interesting to
examine their Penrose limits, following by applying the standard
procedure introduced by \cite{penrose} to the metric and the other
fields of the solution.\footnote{In a string theory context, in
the presence of non-vanishing scalar and tensor fields, the
Penrose limiting procedure for constructing PP-wave solutions was
first applied in \cite{sfepp,gueven}.} The resulting spacetimes
are PP--waves that constitute generalizations of the maximally
supersymmetric PP--wave solution \cite{blaupp} of Type IIB string
theory and, in the context of the AdS/CFT correspondence, are
related to the BMN limit \cite{bmn,sadri} of the gauge theory.
PP--wave limits of marginally-deformed backgrounds were first
considered in \cite{np} and in \cite{LM}, with the former
construction starting from the gauge-theory side of the
correspondence, while further aspects of such solutions were
examined in \cite{pp-deformed}. Here, we further generalize these
constructions to include the effects of $\s$--deformations and of
turning on rotation parameters by following \cite{bs-pp} where
PP--wave solutions based on the solutions of subsection 3.1 were
constructed and further analyzed.

\subsection{Null geodesics in the deformed geometry}

To find the Penrose limits of the deformed solutions, we first need
to identify null geodesics in the respective geometries. The
geodesics we are interested in involve $t$, $r$ and one linear
combination of the cyclic coordinates $\phi_i$ which we denote by
$\phi$, taking the remaining coordinates to constant values
consistent with their equations of motion. To seek such geodesics,
we note that the $\phi_i$ are cyclic and hence setting any of them
to any constant value is automatically consistent, while an ansatz
with constant $\th$ and $\psi$ is certainly consistent if
$\partial_\th G_{ij} \dot{\phi}_i \dot{\phi}_j  =
\partial_\psi G_{ij} \dot{\phi}_i \dot{\phi}_j = 0$ and
$\partial_\th G_{ti} \dot{\phi}_i =
\partial_\psi G_{ti} \dot{\phi}_i = 0$, where the dot denotes
differentiation with respect to the affine parameter $\tau$ of the
geodesic. In our examples with nonzero rotation ($r_0 \ne 0$),
these equations are solved if
\ba
\label{4-1}
&&\th={\pi \ov 2}\ ,\qq \psi,\ \phi_i={\rm any}\ ,\nonumber\\
&&\th=0\ ,\qq \psi=0,{\pi \ov 2}\ ,\qq \phi_i={\rm any}\ ,\\
&&\th=0\ ,\qq \psi={\pi \ov 4}\ ,\qq \phi_2=\phi_3\ ,\qq \phi_1={\rm any}\ .\nonumber
\ea
For the case of zero rotation ($r_0=0$), there emerge additional
solutions, one of which is
\ba
\label{4-2}
&&\th = \arcsin {1 \ov \sqrt{3}}\ ,\qq \psi={\pi \ov 4}\ ,\qq \phi_1 = \phi_2 = \phi_3\ .
\ea
Regarding the sensitivity of the metrics along these geodesics to
$\b$--deformations, we note that the latter are non-trivial only
when $z_1 z_2 + z_2 z_3 + z_3 z_1 \ne 0$ which, by \eqn{3-4},
requires that $\th \ne \pi/2$ and $(\th,\psi) \ne
(0,0),(0,\pi/2)$. Therefore, the effective metrics along the
geodesics in the first two lines of \eqn{4-1} are insensitive to
$\b$--deformations while those along the geodesics in the third
line of \eqn{4-1} and in \eqn{4-2} are sensitive to
$\b$--deformations. Setting the unspecified $\phi_i$ to zero, we
are led to consider the following cases
\ba
\label{4-3}
(J,0,0) :&& \th = {\pi \ov 2}\ ,\quad \psi=0\ ,\quad \phi_1 = \phi ,\quad \phi_2=\phi_3=0\ ,\nonumber\\
(0,J,0) :&& \th = 0\ ,\quad \psi={\pi \ov 2}\ , \quad \phi_2 = \phi\ ,\quad \phi_3=\phi_1=0\ ,\nonumber\\
(0,0,J) :&& \th = 0\ ,\quad \psi=0\ , \quad \phi_3 = \phi ,\quad \phi_1=\phi_2=0\ ,\\
(J,J,J) :&& \th = \arcsin {1 \ov \sqrt{3}}\ ,\quad \psi={\pi \ov 4}\ ,\quad \vphi_3 = \phi\ ,\quad \vphi_1 = \vphi_2 = 0\  ; \quad \hbox{for $r_0 = 0$}\ , \nonumber\\
(0,J,J) :&& \th = 0\ ,\quad \psi={\pi \ov 4}\ ,\quad \vphi \equiv {\phi_2 + \phi_3 \ov 2} = \phi\ ,\quad \chi \equiv {\phi_2 - \phi_3 \ov 2} = 0\ ,\quad \phi_1=0\ , \nonumber
\ea
where, in the fourth line, the $\vphi_i$ are as given in
\eqn{2-11}. The above cases correspond to a particle moving with
angular momenta $(J_{\phi_1},J_{\phi_2},J_{\phi_3})=(J,0,0)$,
$(0,J,0)$, $(0,0,J)$, $(J,J,J)$ and $(0,J,J)$ respectively along
the three isometry directions.

\no
To examine the properties of these geodesics in the various
backgrounds, we distinguish the following cases:
\begin{itemize}
\item {\it Undeformed, zero rotation}. In this case, the
five-sphere is round and the full $SO(6)$ isometry group operates.
All choices correspond to BPS geodesics that can be rotated into
one another. \item {\it Deformed, zero rotation}. Here, the
five-sphere is deformed, with the isometry group broken to
$U(1)^3$. The choices $(J,0,0)$, $(0,J,0)$ and $(0,0,J)$
correspond to three BPS geodesics that can still be rotated into
each other, the choice $(J,J,J)$ corresponds to a distinct BPS
geodesic, and the choice $(0,J,J)$ corresponds to a distinct
non-BPS geodesic. \item {\it Deformed, nonzero rotation}. Now, the
deformed five-sphere is in addition squashed. The available
choices $(J,0,0)$, $(0,J,0)$, $(0,0,J)$ and $(0,J,J)$ are all
generically inequivalent but, for the specific backgrounds
considered in section 3.2, the choices $(0,J,0)$ and $(0,0,J)$
remain equivalent and it suffices to consider only one of them,
say the second.
\end{itemize}
We note that the $(J,J,J)$ geodesic has been first considered for
undeformed ${\rm AdS}_5 \times {\rm S}^5$ in \cite{np} (see also
\cite{pp-deformed}), which is also where Penrose limits of
marginally-deformed ${\rm AdS}_5 \times {\rm S}^5$ first appeared,
found through field-theory considerations.

\subsection{The Penrose limit}

Having identified the geodesics of interest, we are ready to take
the Penrose limit, proceeding along the lines of \cite{bs-pp}. We
first employ the rescaling $(t,\vec{x}_3) \to R^2 ( t,\vec{x}_3)$
and we write the effective metric for $t$, $r$ and $\phi$ as
\be
\label{4-4}
{ds_3^2 \ov R^2} = \g_{tt} dt^2 + \g_{rr} dr^2 + \g_{\phi\phi} d\phi^2 + 2 \g_{t\phi} dt d\phi\ .
\ee
Independence of the metric from $t$ and $\phi$ leads to two
conserved quantities, associated with the Killing vectors
$k=\partial_t$ and $l=\partial_\phi$ and identified with the
energy and the angular momentum, namely
\be
\label{4-5}
E = - k^\m u_\m = - \g_{tt} \dot{t} - \g_{t\phi} \dot{\phi} = 1\ , \qq J = l^\m u_\m = \g_{t\phi} \dot{t} - \g_{\phi\phi} \dot{\phi}\ ,
\ee
Solving for $\dot{t}$ and $\dot{\phi}$ and substituting into
$ds_3^2=0$, we obtain the equation
\be
\label{4-6}
\dot{r}^2 = {\g_{\phi\phi} + 2 J \g_{t\phi} + J^2 \g_{tt} \ov \g_{rr} (\g_{t\phi}^2-\g_{tt} \g_{\phi\phi})}\ ,
\ee
whose solution determines $r$ in terms of $\tau$. We next change
variables from $(r,t,\phi)$ to the new variables $(u,v,y)$,
defined according to
\ba
\label{4-7}
dr &=& \sqrt{{\g_{\phi\phi} + 2 J \g_{t\phi} + J^2 \g_{tt} \ov \g_{rr} (\g_{t\phi}^2-\g_{tt} \g_{\phi\phi})}} du\ , \nonumber\\
dt &=& {\g_{\phi\phi} + J \g_{t\phi} \ov \g_{t\phi}^2-\g_{tt} \g_{\phi\phi}} du - {1 \ov R^2} d v + {J \ov R} d x\ , \\
d\phi &=& - {\g_{t\phi} + J \g_{tt} \ov \g_{t\phi}^2-\g_{tt} \g_{\phi\phi}} du + {1 \ov R} d x\ ,\nonumber
\ea
so that, in particular, $u$ is identified with the affine
parameter $\tau$. We then rescale the spatial brane coordinates as
\be
\label{4-8}
\vec{x}_3 = {\vec{r}_3 \ov R}\ .
\ee
Finally, depending on the case at hand, we make the following
changes of angular variables
\ba
\label{4-9}
(J,0,0) :&& \th = {\pi \ov 2} - {\rho \ov R}\ , \nonumber\\
(0,0,J) :&& \th = {\rho_1 \ov R}\ ,\qq \psi = {\rho_2 \ov R}\ , \\
(J,J,J) :&& \th = \arcsin {1 \ov \sqrt{3}} + {y_1 \ov R}\ ,\; \psi={\pi \ov 4} + {y_2 \ov R}\ ,\; \vphi_1 = - {y_3 -\sqrt{3} y_4 \ov \sqrt{2} R}\ ,\; \vphi_2= {\sqrt{2} y_3 \ov R}\ , \nonumber\\
(0,J,J) :&& \th = {\rho_1 \ov R}\ ,\qq \psi={\pi \ov 4} + {\rho \ov R}\ ,
\qq \chi= {\tilde{\rho} \ov R}\  .
\nonumber
\ea
The Penrose limit is then obtained by substituting all these
changes of variables into the original solution and taking the
limit $R \to \infty$. For the marginally-deformed metrics of
interest, this limit must be taken while keeping $\hgamma \sim R^2
\g$ and $\hsigma \sim R^2 \s$ fixed. The resulting metric always
takes the form of a PP--wave in Rosen-like coordinates and can be
brought to Brinkmann-like coordinates by suitable changes of
variables.

\subsection{PP--wave limits of the deformed solutions}

Here, we employ the method described above to derive the PP--wave
limit of the marginally-deformed metrics of section 3.2. To keep
things relatively simple, we consider only the case $\m=0$,
corresponding to D3-brane distribution, in which case the ``rotation
parameter'' $r_0$ is to be thought of as the radius of the D3-brane
distribution. Note also that, although the $(J,0,0)$ and $(0,0,J)$
geodesics are insensitive to the $\b$--deformation, the PP-wave
backgrounds resulting from the limiting procedure described above
are sensitive to the deformation.

\subsubsection*{The zero-rotation case}

We begin with the zero-rotation case where the deformed metric is
a conformal rescaling of ${\rm AdS}_5 \times {\rm S}^5_\beta$. The
results for the various geodesics are as follows:

\no $\bullet$ {\bf $(J,0,0)$ geodesic.} As a warmup exercise, and
for later reference, we first review the construction of the
simplest PP--wave limit of marginally-deformed ${\rm AdS}_5 \times
{\rm S}^5$, first derived in \cite{LM}, in some detail. The
differential equation for $r$ becomes
\be
\label{4-10}
\dot{r}^2 = 1 - J^2 r^2\ ,
\ee
with solution
\be
\label{4-11}
r^2(u) = {1 \ov J^2} \sin^2 J u\ .
\ee
After following the above limiting procedure we find that
the Penrose limit of the deformed metric reads
\be
\label{4-12}
ds^2 = 2 du dv + A_r d\vec{r}_3^2 + A_x dx^2 + d\vec{y}_4^2 - C du^2\ ,
\ee
where $\vec{y}_4$ is defined by
\be
\label{4-13}
d\vec{y}_4^2 = d\rho^2 + \rho^2 (d\psi^2 + \sin^2 \psi d \phi_2^2 + \cos^2 \psi d \phi_3^2)\ ,
\ee
and the various functions are given by
\be
\label{4-14}
A_r = r^2 = {1 \ov J^2} \sin^2 J u\ ,\qq
A_x = 1- J^2 r^2 = \cos^2 J u\ ,\qq
C = (1+4 |\hbeta|^2) J^2 \vec{y}_4^2\ .
\ee
For future reference, we also write the Rosen form of the
remaining nonzero fields, although we will not do so for the
remaining cases considered below. We have
\ba
\label{4-14a}
\!\!\!\!\!\!B_2 &=& 2 J \hgamma \rho^2 du \wedge (\sin^2 \psi d \phi_2 - \cos^2 \psi d \phi_3)\ ,\nonumber\\
\!\!\!\!\!\!e^{2\Phi} &=& g_{\rm s}^2\, \\
\!\!\!\!\!\!A_2 &=& - {2 J \hsigma \ov g_{\rm s}} \rho^2 du \wedge (\sin^2 \psi d \phi_2 - \cos^2 \psi d \phi_3)\ ,\nonumber\\
\!\!\!\!\!\!A_4 &=& {J \ov g_{\rm s}}  \left( {\sin^4 Ju \ov J^4}
dr_1 \wedge dr_2 \wedge dr_3 \wedge dx + \rho^4 \cos \psi \sin
\psi du \wedge d \psi \wedge d \phi_2 \wedge d \phi_3 \right)\
\nonumber.
\ea
Applying standard transformations\footnote{For a metric in the
Rosen form $ds^2 = 2 du dv + \sum_i A_i (u) dx_i^2 - C du^2$,
applying the sequence of coordinate transformations $x_i \to {x_i
\ov \sqrt{A_i}}$ and $v \to v + {1 \ov 4} \sum_i {d \ln A_i \ov
du} x_i^2$ brings it to the Brinkmann form $ds^2 = 2 dudv + \sum_i
dx_i^2+ ( \sum_i F_i x_i^2 - C ) du^2 $, where $F_i = {1 \ov 4}
\left( {d \ln A_i \ov du} \right)^2 + \ha {d^2 \ln A_i \ov
du^2}$.} to pass to Brinkmann coordinates, we write the metric as
\be
\label{4-15}
ds^2 = 2 du dv + d\vec{r}_4^2 + d\vec{y}_4^2 +( F_r \vec{r}_4^2 + F_y \vec{y}_4^2 ) du^2\ ,
\ee
where $\vec{r}_4 = (\vec{r}_3,x)$ and
\be
\label{4-16}
F_r=-J^2\ ,\qq F_y = - (1+4 |\hbeta|^2) J^2\ ,
\ee
and the remaining nonzero fields of the solution as
\ba
\label{4-17}
H_3 &=& - 4 J \hgamma du \wedge (dy_1 \wedge dy_2 - dy_3 \wedge dy_4)\ , \nonumber\\
e^{2\Phi} &=& g_{\rm s}^2\, \\
F_3 &=& {4 J \hsigma \ov g_{\rm s}}  du \wedge (dy_1 \wedge dy_2 - dy_3 \wedge dy_4)\ , \nonumber\\
F_5 &=& {J \ov g_{\rm s}}  du \wedge (dr_1 \wedge dr_2 \wedge dr_3
\wedge dr_4 - dy_1 \wedge dy_2 \wedge dy_3 \wedge dy_4)\ .
\nonumber
\ea
We see that the deformation affects only the function $F_y$ in the
metric and the NSNS and RR 3-form field strengths.

\no $\bullet$ {\bf $(0,0,J)$ geodesic.} The solution for $r(u)$ is
the same as before and the Penrose limit of the metric in Rosen
coordinates is given by \eqn{4-12} and \eqn{4-14} where
$\vec{y}_4$ is now defined by
\be
\label{4-18}
d\vec{y}_4^2 = (dy_1^2 + dy_2^2) + (dy_3^2 + dy_4^2) = (d\rho_1^2 + \rho_1^2 d \phi_1^2) + (d\rho_2^2 + \rho_2^2 d \phi_2^2)\ ,
\ee
In Brinkmann coordinates, the Penrose limit is given by \eqn{4-16}
and \eqn{4-17}. We explicitly verify that, for $r_0=0$, the
Penrose limits for the $(J,0,0)$ and $(0,J,0)$ geodesics are
equivalent as already remarked in the comments following
\eqn{4-3}.

\no
$\bullet$ {\bf $(J,J,J)$ geodesic.} For this case, it is convenient to introduce
\be
\label{4-19}
\cG^{-1} = 1+{4 |\hbeta|^2 \ov 3}\ ,\qq \cH = 1+{4 \hsigma^2 \ov 3}\ .
\ee
which are just the effective constant values of the functions in
\eqn{3-18} for the given ansatz. Then, the differential equation
for $r$ has the form
\be
\label{4-20}
\cH \dot{r}^2 = 1 - J^2 r^2\ ,
\ee
with solution
\be
\label{4-21}
r^2(u) = {1 \ov J^2} \sin^2 \left( {J  \ov \cH^{1/2}} u \right)\ .
\ee
In Rosen-like coordinates, the Penrose limit of the deformed metric reads
\be
\label{4-22}
ds^2 = 2 du dv + A_r d\vec{r}_3^2 + A_x dx^2 + A_y d\vec{y}_2^2 + \tA_y d \vec{\ty}_2^2 + B_y (y_1 dy_3 - y_2 dy_4) du - C du^2\ ,
\ee
where
\be
\label{4-23}
d\vec{y}_2^2 = dy_1^2 + dy_2^2\ ,\qq d\vec{\ty}_2^2 = dy_3^2 + dy_4^2\ ,
\ee
and
\ba
\label{4-24}
&&A_r = \cH^{1/2} r^2 = {\cH^{1/2} \ov J^2} \sin^2 \left( {J  \ov \cH^{1/2}} u \right)\ ,\nonumber\\
&&A_x = 1 - J^2 r^2 = \cos^2 \left( {J  \ov \cH^{1/2}} u \right)\,\\
&&A_y = \cH^{1/2}\ ,\qq \tA_y = \cG \cH^{1/2}\ ,\qq B_y = - 4 \cG J\ ,\nonumber\\
&&C = {16 J^2 |\hbeta|^2 \cG \vec{y}_2^2 \ov 3 \cH^{1/2}} \ . \nonumber
\ea
In Brinkmann-like coordinates, the metric reads
\be
\label{4-25}
ds^2 = 2 du dv + d\vec{r}_4^2 + d\vec{y}_2^2 + d \vec{\ty}_2^2 + G_2 (y_1 dy_3 - y_2 dy_4) du + ( F_r \vec{r}_4^2 + F_y \vec{y}_2^2 ) du^2\ ,
\ee
where $\vec{r}_4 = (\vec{r}_3,x)$ and
\be
\label{4-26}
F_r = - {J^2 \ov \cH}\ ,\qq F_y= - {16 J^2 |\hbeta|^2 \cG \ov 3 \cH}\ ,\qq G_2 = - {4 \cG^{1/2} \ov \cH^{1/2}}\ ,
\ee
and the remaining fields of the solution are
\ba
\label{4-27}
H_3 &=& - {4 J \ov \sqrt{3} \cH} du \wedge [ 2 \hsigma g_{\rm s} dy_1 \wedge dy_2 + \hgamma \cG^{1/2} (dy_1 \wedge dy_4 + dy_2 \wedge dy_3)]\ , \nonumber\\
e^{2\Phi} &=& \cG \cH^2 g_{\rm s}^2\, \\
F_3 &=& - {4 J \ov \sqrt{3} \cH} du \wedge [ 2 \hgamma dy_1 \wedge dy_2 - {\hsigma \ov g_{\rm s}}  \cG^{1/2} (dy_1 \wedge dy_4 + dy_2 \wedge dy_3)]\ , \nonumber\\
F_5 &=& {J \ov \cH^{3/2} g_{\rm s}}  du \wedge (dr_1 \wedge dr_2
\wedge dr_3 \wedge dr_4 - dy_1 \wedge dy_2 \wedge dy_3 \wedge
dy_4)\ . \nonumber
\ea
This is the generalization of the PP--wave considered in
\cite{np,pp-deformed} which includes the effect of
$\s$--deformations. Now, the deformation affects all the
$F$--functions in the metric and all nonzero fields. This type of
PP--wave falls into the subclass of homogeneous plane waves
considered in \cite{blauloughlin}.

\no $\bullet$ {\bf $(0,J,J)$ geodesic.} For this case, we
introduce
\be
\label{4-28}
\cG^{-1} = 1+|\hbeta|^2\ , \qq \cH = 1+\hsigma^2\ ,
\ee
which are again the effective constant values of the functions in
\eqn{3-18}. Then, the differential equation for $r$ has the form
\be
\label{4-29}
\cH \dot{r}^2 = 1 - {J^2 r^2 \ov \cG}\ ,
\ee
with solution
\be
\label{4-30}
r^2(u) = {\cG \ov J^2} \sin^2 \left( {J \ov \cG^{1/2} \cH^{1/2}} u \right)\ .
\ee
In Rosen-like coordinates, the Penrose limit of the deformed metric reads
\ba
\label{4-31}
&&ds^2 = 2 du dv + A_r d\vec{r}_3^2 + A_x dx^2 + A_y d\vec{y}_2^2 + B_y (y_1 d y_2 - y_2 d y_1) du \nonumber\\ &&+ \:\: A_\rho d\rho^2 + \tA_\rho d \tilde{\rho}^2 + B_\rho \rho d \tilde{\rho} du - C du^2\ ,
\ea
where
\be
\label{4-32}
d\vec{y}_2^2 = d\rho_1^2 + \rho_1^2 d \phi_1^2\ ,
\ee
and
\ba
\label{4-33}
&&A_r = \cH^{1/2} r^2 = {\cG \cH^{1/2} \ov J^2} \sin^2 \left( {J \ov \cG^{1/2} \cH^{1/2}} u \right)\ ,\nonumber\\
&&A_x = \cH^{1/2} \left( \cG - J^2 r^2 \right) = \cG \cH^{1/2} \cos^2 \left( {J \ov \cG^{1/2} \cH^{1/2}} u \right)\ ,\nonumber\\
&&A_y = \cH^{1/2}\ ,\qq B_y = 4 |\hbeta|^2 J\ ,\\
&&A_\rho = \cH^{1/2}\ ,\qq \tA_\rho = \cG \cH^{1/2}\ ,\qq B_\rho = 4 J\ ,\nonumber\\
&&C = {J^2 [ (1-|\hbeta|^2-4|\hbeta|^4) \vec{y}_2^2 - 4 |\hbeta|^2 \rho^2 ] \ov \cH^{1/2}} \ . \nonumber
\ea
In Brinkmann-like coordinates, the metric is given by
\ba
\label{4-34}
&&ds^2 = 2 du dv + d\vec{r}_4^2 + d\vec{y}_2^2 + d\rho^2 + d \tilde{\rho}^2
+ G_y (y_1 d y_2 - y_2 d y_1) du + G_\rho \rho d \tilde{\rho} du \nonumber\\
&& + \:\:(F_r \vec{r}_4^2 + F_y \vec{y}_2^2 + F_\rho \rho^2) du^2\ ,
\ea
where $\vec{r}_4 = (\vec{r}_3,x)$ and
\ba
\label{4-35}
&&F_r= - {J^2 \ov \cG \cH}\ , \nonumber\\
&&F_y= - {J^2 (1-|\hbeta|^2-4|\hbeta|^4) \ov \cH}\ ,\qq G_y = {4 J |\hbeta|^2 \ov \cH^{1/2}}\ , \\
&&F_\rho= {4 J^2 |\hbeta|^2 \ov \cH}\ ,\qq G_\rho = {4 J \ov \cG^{1/2} \cH^{1/2}}\ , \nonumber
\ea
while the remaining fields read
\ba
\label{4-36}
e^{2\Phi} &=& \cG \cH^2 g_{\rm s}^2\, \nonumber\\
F_5 &=& {J \ov \cG^{1/2} \cH^{3/2} g_{\rm s}} du \wedge (dr_1
\wedge dr_2 \wedge dr_3 \wedge dr_4 - dy_1 \wedge dy_2 \wedge
d\rho \wedge d\tilde{\rho})\ .
\ea
We see that only the dilaton and the RR 5-form field strength
survive the Penrose limit along this particular geodesic. The
deformation affects all the $F$--functions in the metric and all
nonzero fields.

\subsubsection*{Two equal rotation parameters (sphere)}

We next consider the case of two equal rotation parameters, where
the new parameter entering the problem is $r_0$. The results for
the various geodesics are as follows:

\no $\bullet$ {\bf $(J,0,0)$ geodesic.} The differential equation
for $r$ becomes
\be
\label{4-37}
\dot{r}^2 = \D_-^2 - J^2 r^2\ ,\qq \D_- \equiv \sqrt{1 - {r_0^2 \ov r^2}}\ .
\ee
Its solution is then given by
\be
\label{4-38}
r^2(u) = {1 \ov 2J^2} ( 1 - a \cos 2Ju )\ , \qq a \equiv \sqrt{1 - 4 J^2 r_0^2}\ ,
\ee
from which it follows that
\be
\label{4-39}
\D_-(u) = \sqrt{1 + {a^2-1 \ov 2( 1 - a \cos 2Ju)}}\ .
\ee
Note that reality requires that, for fixed $r_0$, there is a
maximum angular momentum associated with this trajectory. In Rosen
coordinates, the Penrose limit of the deformed metric reads
\be
\label{4-40}
ds^2 = 2 du dv + A_r d\vec{r}_3^2 + A_x dx^2 + d\vec{y}_4^2 - C du^2\ ,
\ee
where
\be
\label{4-41}
d\vec{y}_4^2 = d\rho^2 + \rho^2 (d\psi^2 + \sin^2 \psi d \phi_2^2 + \cos^2 \psi d \phi_3^2)\ ,
\ee
and
\be
\label{4-42}
A_r = r^2\ ,\qq A_x = \D_-^2 - J^2 r^2\ ,\qq C = (1+4 |\hbeta|^2) J^2 \vec{y}_4^2\ ,
\ee
and are to be understood as functions of $u$ through the
identifications \eqn{4-38} and \eqn{4-39}. In Brinkmann
coordinates, the metric is given by
\be
\label{4-43}
ds^2 = 2 du dv + d\vec{r}_3^2 + dx^2 + d\vec{y}_4^2 + ( F_r \vec{r}_3^2 + F_x x^2 + F_y \vec{y}_4^2 ) du^2\ ,
\ee
with
\ba
\label{4-44}
&&F_r=-J^2 \left[ 1 + {a^2-1 \ov (1-a \cos 2Ju)^2} \right]\ , \nonumber\\
&&F_x=-J^2 \left[ 1 - 3 {a^2-1 \ov (1-a \cos 2Ju)^2} \right]\ , \
\\
&&F_y = - (1+4 |\hbeta|^2) J^2\ .
\nonumber
\ea
Note that, since $0<a<1$ the metric is no-where singular. The
remaining nonzero fields are
\ba
\label{4-45}
H_3 &=& - 4 J \hgamma du \wedge (dy_1 \wedge dy_2 - dy_3 \wedge dy_4)\ , \nonumber\\
e^{2\Phi} &=& g_{\rm s}^2\ , \\
F_3 &=& {4 J \hsigma  \ov g_{\rm s}}  du \wedge (dy_1 \wedge dy_2 - dy_3 \wedge dy_4)\ , \nonumber\\
F_5 &=& {J \ov g_{\rm s}}  du \wedge (dr_1 \wedge dr_2 \wedge dr_3
\wedge dx - dy_1 \wedge dy_2 \wedge dy_3 \wedge dy_4)\ . \nonumber
\ea
These are the same as in Eq. \eqn{4-17} for the same geodesic at
zero rotation. This is in agreement with the results of
\cite{bs-pp}, corresponding to the limiting case $\g=\s=0$, where
it was noted that the RR 5-form field strength at the Penrose
limit for the $(J,0,0)$ geodesic retains the same form as in the
zero-rotation case. Indeed, since the Penrose limit and the
$\b$--deformation procedure commute, the fact that in the
undeformed case the Penrose limits of the metric along the torus
directions, of the dilaton and of the RR 5-form are the same as at
zero rotation guarantees that, in the deformed case, the fields in
\eqn{4-45} will be equal to those in \eqn{4-17}. The deformation
affects only the function $F_y$ in the metric and the NSNS and RR
3-form field strengths while the effect of nonzero rotation
parameters manifests itself only in the functions $F_r$ and $F_x$
in the metric.

\no $\bullet$ {\bf  $(0,0,J)$ geodesic.} The differential equation
for $r$ becomes
\be
\label{4-46}
\dot{r}^2 = 1 - J^2 r^2 \D_-^2\ ,
\ee
with $\D_-$ as in \eqn{4-37}. Its solution is
\be
\label{4-47}
r^2(u) = {b^2 \ov J^2} \sin^2 Ju\ , \qq b \equiv \sqrt{1 + J^2 r_0^2}\ ,
\ee
from which it follows that
\be
\label{4-48}
\D_-(u) = \sqrt{1 - {b^2-1 \ov b^2 \sin^2 Ju}}\ .
\ee
The Penrose limit of the deformed metric reads
\be
\label{4-49}
ds^2 = 2 du dv + A_r d\vec{r}_3^2 + A_x dx^2 + A_y d\vec{y}_2^2 + \tilde{A}_y d\vec{\ty}_2^2 - C du^2\ ,
\ee
where
\be
\label{4-50}
d\vec{y}_2^2 = dy_1^2 + dy_2^2 = d\rho_1^2 + \rho_1^2 d \phi_1^2\ ,\qq d\vec{\ty}_2^2 = dy_3^2 + dy_4^2 = d\rho_2^2 + \rho_2^2 d \phi_2^2\ ,
\ee
and
\ba
\label{4-51}
&&A_r = r^2 \D_-\ ,\qq A_x = {1 \ov \D_-} - J^2 r^2 \D_-\ ,\qq A_y = \D_-\ ,
\qq \tilde{A}_y = {1 \ov \D_-}\ ,\nonumber\\
&&C = J^2 \left[ \left( {\vec{y}_2^2 \ov \D_-} + \vec{\ty}_2^2 \D_- \right)
+ 4 |\hbeta|^2 \left( \vec{y}_2^2 \D_- + {\vec{\ty}_2 \ov \D_-} \right) \right]\ , \nonumber
\ea
and again are to be understood as functions of $u$ through
\eqn{4-46} and \eqn{4-47}. In Brinkmann coordinates, the metric
reads
\be
\label{4-52}
ds^2 = 2 du dv + d\vec{r}_3^2 + dx^2 + d\vec{y}_2^2 + d\vec{\tilde{y}}_2^2 + (F_r \vec{r}_3^2 + F_x x^2 + F_y \vec{y}_2^2 + \tilde{F}_y \vec{\tilde{y}}_2^2) du^2\ ,
\ee
with
\ba
\label{4-53}
&& F_r = -J^2 \left[1 +  {b^2-1\ov 4 (1-b^2 \cos^2 J u)^2}\left({b^2-1\ov \sin^2Ju} -b^2+3\right) \right]\ ,
\nonumber\\
&& F_x = -J^2 \left[ 1- {5\ov 4} {b^2-1\ov (1-b^2 \cos^2 J u)^2} -{1\ov 4} {b^2-1\ov \sin^2 Ju (1-b^2 \cos^2 J u)}\right]\ ,
\\
&& F_y= -J^2 \left[{b^2 \sin^2Ju\ov 1-b^2 \cos^2 Ju} - {(b^2-1)(4 b^2\cos^4 Ju-2-(b^2+1)\cos^2 Ju) \ov 4
(1-b^2 \cos^2Ju )^2 \sin^2 Ju} + 4 |\hbeta|^2 \right] \ ,
\nonumber\\
&& \tilde F_y = -J^2 \left[{1-b^2 \cos^2 Ju\ov b^2 \sin^2Ju} + {(b^2-1)(4 b^2 \cos^4 Ju-2+(1-3 b^2)\cos^2 Ju)\ov
4(1-b^2 \cos^2Ju )^2 \sin^2 Ju } + 4 |\hbeta|^2 \right]\ ,
\nonumber
\ea
while the remaining nonzero fields are
\ba
\label{4-54}
H_3 &=& - 4 J \hgamma du \wedge (dy_1 \wedge dy_2 - dy_3 \wedge dy_4)\ , \nonumber\\
e^{2\Phi} &=& g_{\rm s}^2\ , \\
F_3 &=& {4 J \hsigma \ov g_{\rm s}}  du \wedge (dy_1 \wedge dy_2 - dy_3 \wedge dy_4)\ , \nonumber\\
F_5 &=& {J \ov 2 g_{\rm s}} {1-b^2 \cos 2Ju \ov b \sin Ju
\sqrt{1-b^2 \cos^2 Ju}} du \wedge (dr_1 \wedge dr_2 \wedge dr_3
\wedge dx - dy_1 \wedge dy_2 \wedge dy_3 \wedge dy_4)\ . \nonumber
\ea
In contrast to the previous case, these differ from the
corresponding expression for the same geodesic at zero rotation in
that the RR 5-form now has an $r_0$--dependent overall
coefficient. Again this is in agreement with the corresponding
analysis of \cite{bs-pp} for the limiting case $\g=\s=0$. Now, the
deformation affects only the functions $F_y$ and $\tilde{F}_y$ in
the metric and the NSNS and RR 3-form field strengths while the
effect of turning on rotation parameters manifests itself in all
$F$--functions in the metric and in the RR 5-form field strength.

\no $\bullet$ {\bf $(0,J,J)$ geodesic.} Now, it is convenient to
introduce
\be
\label{4-55}
\cG^{-1} =  {(1+|\hbeta|^2) r^2 - r_0^2 \ov r^2 - r_0^2}\ , \qq \cH = { (1+\hsigma^2) r^2 - r_0^2 \ov r^2 - r_0^2}\ ,
\ee
which are the effective values of the functions in \eqn{3-24} for
the given ansatz, now being functions of $r$. The differential
equation for $r$ has the form
\be
\label{4-56}
\cH \dot{r}^2 = 1 - {J^2 r^2 \D_-^2 \ov \cG}\ ,
\ee
with $\D_-$ as in \eqn{4-37}. Unfortunately, this equation cannot
be solved explicitly in the general case, for which we will
content ourselves with presenting the Penrose limit only in the
Rosen-like form, with the $u$--dependence of the metric components
entering implicitly through \eqn{4-54}. This metric is given by
\ba
\label{4-57}
&&ds^2 = 2 du dv + A_r d\vec{r}_3^2 + A_x dx^2 + A_y d\vec{y}_2^2 + B_y (y_1 dy_2 - y_2 dy_1) du \nonumber\\ && + \:\: A_\rho d\rho^2 + \tA_\rho d \tilde{\rho}^2 + B_\rho \rho d \tilde{\rho} du - C du^2\ ,
\ea
where
\be
\label{4-58}
d\vec{y}_2^2 = d\rho_1^2 + \rho_1^2 d \phi_1^2\ ,
\ee
and
\ba
\label{4-59}
&&A_r = \cH^{1/2} r^2 \D_-\ ,\qq A_x = \cH^{1/2} \left( {\cG \ov \D_-} - J^2 r^2 \D_- \right)\ ,\nonumber\\
&&A_y = \cH^{1/2} \D_-\ ,\qq B_2 = 4 |\hbeta|^2 J\ ,\\
&&A_\rho = {\cH^{1/2} \ov \D_-}\ ,\qq \tA_w = {\cG \cH^{1/2} \ov \D}\ ,\qq B_\rho = 4 J\ ,\nonumber\\
&&C = {J^2 [ (1-|\hbeta|^2-4|\hbeta|^4) \vec{y}_2^2 - 4 |\hbeta|^2 \rho^2 ] \ov \cH^{1/2} \D_-} \ . \nonumber
\ea
On the other hand, for the case of a pure $\g$--deformation, Eq.
\eqn{4-56} {\em can} be solved exactly with the result
\be
\label{4-60}
r^2(u) = {b^2 \cG_0 \ov J^2} \sin^2 \left( { J u \ov \cG_0^{1/2} }\right)\ ,
\ee
where $b$ is as in \eqn{4-47} and
\be
\label{4-61}
\cG_0^{-1} = 1 + \hgamma^2\ .
\ee
The various functions for the metric in Rosen-like coordinates are
found by substituting \eqn{4-60} into \eqn{4-57} and setting $\cG
\to \cG_0$ and $\cH \to 1$. In Brinkmann-like coordinates, the
formulas for the various functions are rather messy and we refrain
from quoting them. We just note that the solution has $F_1=F_3=0$
and that all $F$--functions in the metric and all nonzero fields
are affected both by the deformation and by the rotation
parameters.

\subsubsection*{One rotation parameter (disc)}

We finally consider the case of one rotation parameter. As
remarked earlier on, when $\m=0$, the solutions for one rotation
parameter are obtained from those for two equal rotation
parameters through the replacement $r_0^2 \to -r_0^2$. Making this
replacement in the corresponding Penrose limits, we find the
following results:

\no $\bullet$ {\bf $(J,0,0)$ geodesic.} The Penrose limit of the
solution in Brinkmann coordinates is given by Eqs.
\eqn{4-43}--\eqn{4-45}, now with
\be
\label{4-62} a = \sqrt{1 + 4 J^2 r_0^2}\ .
\ee
Note that now $a$ is real for all values of $J r_0$. Also, since
$a>1$ the PP-wave metric is singular at $\cos 2 J u = 1/a$.

\no $\bullet$ {\bf  $(0,0,J)$ geodesic.} The Penrose limit of the
solution in Brinkmann coordinates is given by Eqs.
\eqn{4-52}--\eqn{4-54}, now with
\be
\label{4-63} b = \sqrt{1 - J^2 r_0^2}\ .
\ee

\no $\bullet$ {\bf $(0,J,J)$ geodesic.} In the general case, the
Penrose limit of the deformed metric in Rosen-like coordinates is
given by Eq. \eqn{4-57} with $d\vec{y}_2^2$ given in \eqn{4-58}
and the $A$--, $B$-- and $C$-- functions given in \eqn{4-59} but
with $\D_-$ replaced by $\D_+=\sqrt{1 + {r_0^2 \ov r^2}}$ and with
$\cG^{-1}$ and $\cH$ appropriately modified. For the case of a
pure $\g$--deformation, there exists the explicit solution
\eqn{4-58} for $r(u)$ and the Penrose limit in Rosen-like
coordinates is found by substituting that solution into Eqs.
\eqn{4-57}-\eqn{4-59}, with $\D_-$ replaced by $\D_+$.

\section{Giant gravitons on $\b$--deformed PP--waves}
\label{sec5}

Given the marginally-deformed geometries, it is interesting to
investigate the various extended objects that they can support. In
this respect, an important role is played by BPS configurations of
spherical D3-branes, the so-called giant gravitons. To summarize
the basic facts, the authors of \cite{mcgreevy}, building on
results of \cite{myers}, considered a KK excitation (graviton) in
${\rm AdS}_5 \times {\rm S}^5$ with nonzero angular momentum along
an ${\rm S}^5$ direction and contemplated the possibility that it
might blow up on an ${\rm S}^3$ inside the ${\rm S}^5$ without
raising its energy. They found that such a state (the giant
graviton) can indeed exist, with the blowing up being due to its
angular momentum and the extra force required to keep it stable
under shrinking being provided by RR repulsion. Soon after that,
it was found \cite{dual-giants} that there also exist ``dual''
giant gravitons with similar properties, supported on the ${\rm
S}^3$ inside the ${\rm AdS}_5$ part of the geometry. The
construction of giant gravitons has been extended towards various
directions in \cite{giant-more}, while investigations from the
dual field-theory side have been carried out in \cite{giant-sym}.
An interesting result is that, in certain cases
\cite{squashed-giants,unstable-giants} where the geometry of the
${\rm S}^5$ supporting the giant graviton is deformed, the latter
has higher energy than the point graviton, and may even not exist
at all as a solution.

\no This latter fact serves as a motivation for examining giant
gravitons in the marginally-deformed solutions of interest, since
one of the main features of the latter is precisely a deformation
of ${\rm S}^5$. In the existing literature, giant gravitons have
been considered only for the case of $\g$--deformations. For the
full deformed solutions, giant gravitons were
first constructed in \cite{unstable-giants} for a special case of the
non-supersymmetric three-parameter background of
\cite{frolov,frt} and, more recently, for the general three-parameter \cite{pirrone} and the single
parameter \cite{pirrone,imeroni} backgrounds; in particular, the giant
gravitons of \cite{pirrone,imeroni} were found to be independent of
the deformation parameter and hence still degenerate with the
point graviton. For the PP-wave limits of the marginally-deformed
backgrounds, giant gravitons have been constructed, in analogy to
the considerations of \cite{giant-pp} for the PP--wave limits of
${\rm AdS}_5 \times {\rm S}^5$, in \cite{hamilton} for the two PP--wave limits of $\g$--deformed
${\rm AdS}_5 \times {\rm S}^5$, namely those along the $(J,0,0)$
and the $(J,J,J)$ geodesic. For the first geodesic, the solution was
constructed exactly and it was found that $\g$--deformations do
not lift the degeneracy of the giant and point gravitons, in
accordance with \cite{imeroni}, and a stability analysis indicated
that the former is perturbatively stable. For the second geodesic, the
problem was attacked perturbatively in $\hgamma$, with the
leading-order analysis indicating that the $\g$--deformation lifts
the degeneracy in favor of the point graviton, but no definitive
conclusion on whether the giant graviton survives the deformation
for large enough $\hgamma$ was reached.

\no Here, we address the question of identifying giant gravitons
on the PP--wave limit of the deformed geometries, this time in the
presence of $\s$--deformations and/or nonzero rotation parameters.
Restricting to PP--waves along the $(J,0,0)$ geodesic, our exact
analysis for the giant graviton residing on the deformed ${\rm
S}^5$ part of the geometry shows that $\s$--deformations have an
altogether different effect than that of $\g$--deformations,
lifting the degeneracy of the giant and point gravitons and, for
$\hsigma$ above a critical value, completely removing the giant
graviton from the spectrum. Moreover, the analysis of small
fluctuations of the giant graviton reveals that the latter is
perturbatively stable throughout its range of existence. We also
consider dual giant gravitons residing on the ${\rm AdS}_5$ part
of the geometry, in which case the deformation does not affect
neither the solution, nor its stability properties. In what
follows, we present the relevant analysis, keeping a similar
notation with \cite{hamilton} in order to facilitate comparison.

\subsection{Giant gravitons on the deformed PP--waves}

To describe the giant-graviton solutions in the PP--wave
spacetimes of interest, we consider the action for a probe
D3-brane in this background, given by the sum of the
Dirac--Born--Infeld and Wess--Zumino terms,
\be
\label{5-4} S_{{\rm D}3} = S_{\rm DBI} + S_{\rm WZ} = - T_3 \int
d^4 \s e^{-\Phi} \sqrt{ - \det \cP[G-B] } + T_3 \int \sum_q \cP[
A_q \wedge e^{-B_2} ]\ ,
\ee
where $T_3$ is the D3-brane tension, equal to $1 / (2\pi)^3$ in
units where $\alpha^{\prime}=1$, and $\cP[f]$ stands for the
pullback of a spacetime field $f$ on the worldvolume. Below we
describe the construction of giant gravitons and dual giant
gravitons on the deformed PP--waves under consideration.

\subsubsection{Giant gravitons}

Starting with ordinary giant gravitons, we want to describe
a D3-brane wrapping the ${\rm S}^3$ inside the ${\rm S}^5_\b$
in the deformed geometry. To do so, we employ the gauge choice
\be
\label{5-5} \tau = u\ ,\qq \sigma_1 = \psi\ ,\qq \sigma_2 =
\phi_2\ ,\qq \sigma_3 = \phi_3\ ,
\ee
and we consider the ansatz
\be
\label{5-6} v = - \n u\ ,\qq \vec{r}_4 = 0\ ,\qq
\rho=\rho_0=\const\ .
\ee
Noting that the spatial brane coordinates are just the angular
coordinates employed in the Rosen form \eqn{4-12}--\eqn{4-14a} of
the PP--wave solution of interest, the pullbacks of the various
fields on the D3-brane can be immediately read off from these
equations. Setting for convenience $J=1$, we find
\ba
\label{5-8} \cP[G] = \diag \left( - 2 \n - (1+4 |\hbeta|^2)
\rho_0^2, \, \rho_0^2 , \, \rho_0^2 \sin^2 \psi , \, \rho_0^2
\cos^2 \psi \right)\ ,
\ea
and
\ba
\label{5-9}
\cP[B_2] &=& 2 \hgamma \rho_0^2 d u \wedge (\sin^2 \psi d \s_2 - \cos^2 \psi d \s_3)\ ,\nonumber\\
\cP[A_2] &=& - {2 \hsigma \ov g_{\rm s}} \rho_0^2 d u \wedge (\sin^2 \psi d \s_2 - \cos^2 \psi d \s_3)\ ,\\
\cP[A_4] &=& {1 \ov g_{\rm s}} \rho_0^4 \cos \psi \sin \psi d u
\wedge d \s_1 \wedge d \s_2 \wedge d \s_3\ .\nonumber
\ea
From these equations, we readily compute
\be
\label{5-10} - \det \cP[G-B] =  [ 2 \n + (1+4 \hsigma^2) \rho_0^2]
\rho_0^6 \cos^2\psi \sin^2\psi\ .
\ee
As in \cite{hamilton}, we note that all $\hgamma$ dependence has
dropped out due to the cancellation of terms coming from the
metric and from $B_2$. The important fact is that, since the
metric now involves $|\hbeta|^2$ in place of $\hgamma^2$ while
$B_2$ still depends only on $\hgamma$, there remains a non-trivial
dependence on the deformation through the parameter $\hsigma$.
Noting also that, since $\cP[B_2 \wedge A_2] = 0$, only $\cP[A_4]$
contributes to the Wess--Zumino action, we write the full D3-brane
action as
\ba
\label{5-11} S_{{\rm D}3} &=& - {T_3 \ov g_{\rm s}} \int d u
d\Omega_3 [ \rho_0^3 \sqrt{2 \n + (1+4 \hsigma^2)  \rho_0^2} -
\rho_0^4 ] = \int d u L_{{\rm D}3}\ ,
\ea
where $d\Omega_3 = \cos\psi \sin\psi d\psi d\phi_1 d\phi_2$ is the
${\rm S}^3$ volume element and $L_{{\rm D}3}$ is the Lagrangian
\be
\label{5-12} L_{{\rm D}3} = -M [ \rho_0^3 \sqrt{2\n + (1+4
\hsigma^2) \rho_0^2} - \rho_0^4] \ ,
\ee
where
\be
\label{5-13} M = {2\pi^2 T_3 \ov g_{\rm s}} = {1 \ov 4 \pi g_{\rm
s}} = {N \ov R^4}\ .
\ee
Since $L_{{\rm D}3}$ is independent of $v$ and $u$, we have two
conserved first integrals, given by the light-cone momentum
\be
\label{5-14} P = - {\partial L_{{\rm D}3} \ov \partial \n} = {M
\rho_0^3 \ov \sqrt{2\n + (1+4 \hsigma^2) \rho_0^2} }\ ,
\ee
and the light-cone Hamiltonian
\be
\label{5-15} E = \n {\partial L_{{\rm D}3} \ov \partial \n} -
L_{{\rm D}3} = {M \rho_0^3 [ \n + (1+4 \hsigma^2) \rho_0^2] \ov
\sqrt{2 + (1+4 \hsigma^2) \rho_0^2} } - M \rho_0^4\ .
\ee
Solving \eqn{5-14} for $\n$ and substituting in \eqn{5-15}, we
write
\be
\label{5-16} E = {M^2 \ov 2 P} \rho_0^6  - M \rho_0^4 + {(1+4
\hsigma^2) P \ov 2} \rho_0^2\ .
\ee
As a function of $\rho_0$, $E$ has local extrema at the radii
\be
\label{5-17} \rho_0 = 0 ,\qq \rho_0 = \rho_{0\pm} = \sqrt{(2 \pm
\D) P \ov 3 M}\ ,
\ee
where we have defined
\be
\label{5-18}
\D \equiv \sqrt{1 - 12\hsigma^2}\ .
\ee
The corresponding light-cone energies are given by
\be
\label{5-19} E_0 = 0\ ,\qq E_{\pm} = (2 \pm \D)^2 (1 \mp \D) {P^2
\ov 27 M}\ .
\ee
Note also that \eqn{5-14} and the second of \eqn{5-17} lead to the
constraint
\be
\label{5-19a} \nu = - {2 \ov 9} \rho_{0\pm}^2 \left(2 \pm
\D\right) \left(1 \mp \D\right)\ ,
\ee
determining $\nu$ in terms of $\rho_{0\pm}$.

\begin{figure}[!t]
\begin{center}
\begin{tabular}{c}
\includegraphics [height=6cm]{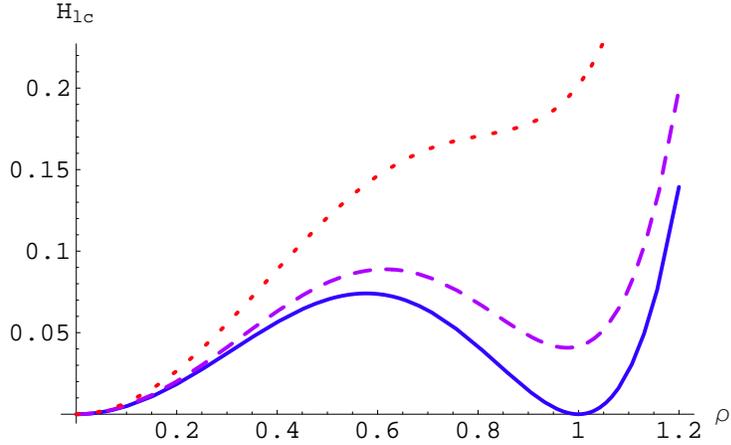}
\end{tabular}
\end{center}
\caption{Light-cone Hamiltonian for the PP--wave along the
$(J,0,0)$ geodesic plotted as a function of $\rho$. The three
curves correspond to $\hsigma=0$ (solid), $\hsigma={0.5 \ov 2
\sqrt{3}}$ (dashed) and $\hsigma={1.1 \ov 2 \sqrt{3}}$ (dotted).
The plots are shown in units where $M=P=1$} \label{fig1}
\end{figure}

\no For $0 \leqslant \hsigma < {1 \ov 2 \sqrt{3}}$, the radius
$\rho_0=\rho_{0-}$ corresponds to a local maximum, while the radii
$\rho_0 = 0$ and $\rho_0 = \rho_{0+}$ correspond to two local
minima, the point graviton and the giant graviton. At exactly
$\hsigma = 0$, we recover the usual result that $E_+ = E_0$, i.e.
that the giant graviton is degenerate in energy with the point
graviton. However, for $0 < \hsigma < {1 \ov 2 \sqrt{3}}$, we have
$E_+ > E_0$ i.e. the degeneracy is lifted with the giant graviton
becoming energetically unfavorable. At $\hsigma = {1 \ov 2
\sqrt{3}}$, the radii $\rho_0 = \rho_{0-}$ and $\rho_0 = \rho_{0+}$
degenerate into a saddle point, leaving only one minimum at
$\rho_0=0$. Finally, for $\hsigma > {1 \ov 2 \sqrt{3}}$, the only
extremum is the minimum at $\rho_0 = 0$: the giant graviton
disappears from the spectrum. The situation is depicted in Fig. 1.
To summarize, for the PP--wave along the $(J,0,0)$ geodesic, complex
$\b$--deformations, unlike $\g$--deformations, have the effect of
lifting the degeneracy of the giant and point gravitons for small
values of $\hsigma$ and of removing the giant graviton from the
spectrum for large values of $\hsigma$. Also, as we shall see
explicitly later on, although the effective Lagrangian \eqn{5-12}
depends only on $\hsigma$, the spectrum of small perturbations about
the giant graviton solution is dependent on both deformation
parameters.

\no We note that the above results remain valid when the rotation
parameter $r_0$ is turned on. This follows from the fact that the
relevant components of the metric and the remaining fields are
identical to those at zero rotation (see the comments following
Eq. \eqn{4-45}).

\subsubsection{Dual giant gravitons}

Proceeding to the case of dual giant gravitons, we have to
consider a D3-brane wrapping the ${\rm S}^3$ originating from the
${\rm AdS}_5$ part of the geometry. Since the latter part of the
geometry is unaffected by the deformation, it is immediately seen
that the dual giant graviton solution exists and is independent of
the deformation; however, to set up the notation for the stability
analysis that follows, let us demonstrate it explicitly. To do so,
we first parametrize, in analogy to \eqn{4-13}, the coordinate
vector $\vec{r}_4$ by the coordinates
$(\tilde{\rho},\tilde{\psi},\tilde{\phi}_2,\tilde{\phi}_3)$
defined by
\be
\label{5-a1} d\vec{r}_4^2 = d \tilde{\rho}^2 + \tilde{\rho}^2
(d\tilde{\psi}^2 + \sin^2 \tilde{\psi} d \tilde{\phi}_2^2 + \cos^2
\tilde{\psi} d \tilde{\phi}_3^2)\ ,
\ee
we employ the gauge choice
\be
\label{5-a2} \tau = u\ ,\qq \sigma_1 = \tilde{\psi}\ ,\qq \sigma_2
= \tilde{\phi}_2\ ,\qq \sigma_3 = \tilde{\phi}_3\ ,
\ee
and we consider the ansatz
\be
\label{5-a3} v = - \n u\ ,\qq \vec{y}_4 = 0\ ,\qq
\tilde{\rho}=\tilde{\rho}_0=\const\ .
\ee
Proceeding as before, we find that the only relevant pullbacks of
the spacetime fields are
\ba
\label{5-a4} \cP[G] = \diag \left( - 2 \n - \tilde{\rho}_0^2, \,
\tilde{\rho}_0^2 , \, \tilde{\rho}_0^2 \sin^2 \tilde{\psi} , \,
\tilde{\rho}_0^2 \cos^2 \tilde{\psi} \right)\ ,
\ea
and
\ba
\label{5-a5} \cP[A_4] &=& {1 \ov g_{\rm s}} \tilde{\rho}_0^4 \cos
\tilde{\psi} \sin \tilde{\psi} d u \wedge d \s_1 \wedge d \s_2
\wedge d \s_3\ ,
\ea
and we arrive at the action
\be
\label{5-a6} S_{{\rm D}3} = \int d u L_{{\rm D}3}\ , \qq L_{{\rm
D}3} = -M \left( \tilde{\rho}_0^3 \sqrt{2\n + \tilde{\rho}_0^2} -
\tilde{\rho}_0^4 \right) \ ,
\ee
with $M$ as in \eqn{5-13}. The dual giant graviton solution is
found as before and is obviously independent of the deformation,
which implies that it is degenerate with the point graviton for
all values of the deformation parameters. However, as we shall
see, this degeneracy does not extend to the spectrum of
fluctuations around this solution.

\no When the rotation parameter $r_0$ is turned on, however, the
above solution is no longer valid. This is because the $SO(4)$
symmetry of the $\vec{r}_4 = (\vec{r}_3,x)$ directions is broken
by the rotation parameter, as manifested by the fact that the
functions $F_r(u)$ and $F_x(u)$ in \eqn{4-43} are different.

\subsection{Small fluctuations and perturbative stability}

We now turn to an analysis of small (bosonic) fluctuations about the
giant graviton configurations in deformed PP--waves, following the
treatment of \cite{giant-vibrations}. We consider both ordinary and
dual giant gravitons and, for the former case, we also take account
of the presence of rotation. As we shall see, although the spectrum
of fluctuations is affected by the deformation, the standard result
that these configurations are perturbatively stable remains
unchanged.

\subsubsection{Giant gravitons}

Starting with ordinary giant gravitons, the perturbation of the
classical configuration is described by keeping the gauge choice
as in \eqn{5-5} and perturbing the embedding as
\ba
\label{5-21} & v = - \n u + \delta v(u,\psi,\phi_2,\phi_3)\ , \qq
\rho = \rho_0 + \delta \rho(u,\psi,\phi_2,\phi_3)\ , \nonumber\\
&\vec{r}_3 = \d\vec{r}_3 (u,\psi,\phi_2,\phi_3)\ , \qq x = \d x
(u,\psi,\phi_2,\phi_3)\ .
\ea
Computing the pullbacks of the various fields as before, inserting
them in the D3-brane action and expanding up to second order in
the fluctuations, we write
\be
\label{5-22}
S_{{\rm D}3} = S_0 + S_1 + S_2 + \ldots\ ,
\ee
where the various terms correspond to the respective powers of the
fluctuations. The zeroth-order is just the classical action given
by \eqn{5-11} and \eqn{5-12}. The linear term reads
\ba
\label{5-24}
\!\!\!\!\!\!\!\!\!\!\!\!\!\!\!\!S_1 &=& -{M \rho_0^2
\ov 2 \pi^2 \sqrt{2 \n + (1+4\hsigma^2) \rho_0^2}} \nonumber\\
\!\!\!\!\!\!\!\!\!\!\!\!\!\!\!\!&\times& \int d u d\Omega_3
\left\{ \rho_0 \partial_u \d v - 2 \left[ 3\n + 2
(1+4\hsigma^2)\rho_0^2 - 2 \rho_0\sqrt{2\n +
(1+4\hsigma^2)\rho_0^2} \right] \d \rho \right\}\ ,
\ea
and it depends only on the deformation parameter $\hsigma$. The
first term is a total derivative that vanishes upon integration
over $u$, while the second term vanishes upon imposing the
constraint \eqn{5-19a}. Finally, calculating the quadratic term
and making use of \eqn{5-19a} (with the upper signs), we obtain
\ba
\label{5-26} S_2 &=& - {3 M \rho_0^2 \ov 4 \pi^2 (2 + \D)} \int du
d\Omega_3 \biggl\{ {8 \D(1-\D) \ov 3} \d \rho^2 - F_r(u)
\d\vec{r}_3^2 - F_x(u) \d x^2 \nonumber\\
&& \qq\qq\qq\qq - {12 \D \ov (2 + \D) \rho_0} \d \rho \partial_u
\d v + 4 \hgamma^2 [ (\partial_{\phi_2}
- \partial_{\phi_3}) \d\vec{r}_5]^2 \nonumber\\
&& \qq\qq\qq\qq - {9 \ov (2+\D)^2 \rho_0^2} (\partial_u \d v)^2 -
(\partial_u \d \vec{r}_5)^2
\\
&& \qq\qq\qq\qq + {1 \ov \rho_0^2} h^{\a\b} \partial_\a \d v
\partial_\b \d v
+ {(2+\D)^2 \ov 9} h^{\a\b} \partial_\a \d \vec{r}_5 \cdot
\partial_\b \d \vec{r}_5 \biggr\}\ .
\nonumber
\ea
Here, $\D$ is the $\hsigma$--dependent quantity defined in
\eqn{5-18}, $h_{\alpha\beta}$ is the metric on ${\rm S}^3$ and we
introduced the shorthand $\d \vec{r}_5 = (\d \vec{r_3},\d
x,\d\rho)$. Also, $F_r(u)$ and $F_x(u)$ are the $u$--dependent
functions defined in \eqn{4-44}, with the parameter $a$ given by
the second of \eqn{4-38} and by \eqn{4-62} for the case of two and
one rotation parameters respectively.  In the limit of zero
rotation both $F_r$ and $F_x$ equal to $-1$. We note that, unlike the zeroth-order
and linear terms, the quadratic term has some, {\it albeit}
restricted, dependence on the deformation parameter $\hgamma$ in
addition to the dependence on $\hsigma$. After several
integrations by parts, the quadratic action becomes
\ba
\label{5-27} S_2 &=& - {3 M \rho_0^2 \ov 4 \pi^2 (2 + \D)}\int du
d\Omega_3 \biggl\{ {8 \D(1-\D) \ov 3} \d \rho^2
- F_r(u) \d\vec{r}_3^2 - F_x(u) \d x^2 \nonumber\\
&& \qq\qq\qq - {12 \D \ov (2 + \D) \rho_0} \d \rho \partial_u \d v
+ {1 \ov \rho_0^2} \d v \left[ \left(3 \ov
2+\D\right)^2 \partial_u^2 - \D_{{\rm S}^3} \right] \d v
\\
&& \qq\qq\qq + \d \vec{r}_5 \cdot \left[
\partial_u^2  - \left( {2+\D \ov 3} \right)^2 \D_{{\rm S}^3}
- 4 \hgamma^2 (\partial_{\phi_2} -
\partial_{\phi_3})^2 \right] \d\vec{r}_5 \biggr\}\ ,
\nonumber
\ea
where $\D_{{\rm S}^3}$ is the Laplacian on ${\rm S}^3$. To
proceed, we may expand all fluctuations in the basis spanned by
the combinations $\Psi_{{\rm S}^3,\ell n_2
n_3}(\psi,\phi_2,\phi_3)$ of ${\rm S}^3$ harmonics having definite
quantum numbers under the two $U(1)$'s corresponding to shifts of
$\phi_2$ and $\phi_3$, i.e. by the simultaneous eigenfunctions of
the operators $\D_{{\rm S}^3}$ and $\partial_{\phi_{2,3}}$ with
\ba
\label{5-29}
\D_{{\rm S}^3} \Psi_{{\rm S}^3,\ell n_2 n_3} (\psi,\phi_2,\phi_3)
&=& - \ell (\ell+2) \Psi_{{\rm S}^3,\ell n_2 n_3} (\psi,\phi_2,\phi_3)\ ,
\nonumber\\
\partial_{\phi_{2,3}} \Psi_{{\rm S}^3,\ell n_2 n_3} (\psi,\phi_2,\phi_3)
&=& {\rm i} n_{2,3} \Psi_{{\rm S}^3,\ell n_2 n_3}
(\psi,\phi_2,\phi_3)\ ,
\ea
where $\ell$, $n_2$ and $n_3$ are required to satisfy\footnote{The explicit expression for the
$\Psi_{{\rm S}^3,\ell n_2 n_3}$'s can be written in terms of Jacobi Polynomials (see, for instance,
section 6 of the first of \cite{hsz}).}
\be
\ell=0,1,\ldots\ ,\qq \ell - |n_2| - |n_3| = 2 k\ ,\quad
k=0,1,\ldots\ .
\ee
For the $\d v$ and $\d \rho$ fluctuations, the fact that their
coefficients in the action are $u$--independent allows us to
introduce an $e^{-{\rm i} \omega u}$ time dependence and thus we
can write
\ba
\label{5-28} \d v (u,\psi,\phi_2,\phi_3) &=& \d v_{\ell n_2 n_3}
e^{-{\rm i} \omega u} \Psi_{{\rm S}^3,\ell n_2 n_3}
(\psi,\phi_2,\phi_3)\ ,\nonumber\\
\d \rho (u,\psi,\phi_2,\phi_3) &=& \d \rho_{\ell n_2 n_3} e^{-{\rm
i} \omega u} \Psi_{{\rm S}^3,\ell n_2 n_3} (\psi,\phi_2,\phi_3)\ ,
\ea
while for the $(\d \vec{r}_3,\d x)$ fluctuations, we can only set
\ba
\label{5-28a} \d \vec{r}_3 (u,\psi,\phi_2,\phi_3) &=& \d
\vec{r}_{3,\ell n_2 n_3}(u) \Psi_{{\rm S}^3,\ell n_2 n_3}
(\psi,\phi_2,\phi_3)\
,\nonumber\\
\d x (u,\psi,\phi_2,\phi_3) &=& \d x_{\ell n_2 n_3}(u) \Psi_{{\rm
S}^3,\ell n_2 n_3} (\psi,\phi_2,\phi_3)\ .
\ea

\no The spectrum of fluctuations follows from inserting the
expansions \eqn{5-28} and \eqn{5-28a} in the equations of motion
stemming from the action \eqn{5-27}. Starting from the null
fluctuation $\d v$ and the radial fluctuation $\d \rho$, we find
that they satisfy the coupled system
\be
\label{5-31} \left(
\begin{array}{cc}
{(2+\D)^2 \ell(\ell+2) - 9 \omega^2 \ov (2+\D)^2 \rho_0^2} & - {6 {\rm i} \D \omega \ov (2+\D) \rho_0} \\
{6 {\rm i} \D \omega \ov (2+\D) \rho_0} & {(2+\D)^2 \ell(\ell+2) -
9 \omega^2 + 36 \hgamma^2 (n_2-n_3)^2 + 24 \D(1-\D) \ov 9}
\end{array}
\right) \left( \begin{array}{c} \d v_{\ell n_2 n_3} \\ \d
\rho_{\ell  n_2 n_3}
\end{array} \right) = 0\ ,
\ee
which leads to the following spectrum
\ba
\label{5-33}
\omega_{\pm,\ell n_2 n_3}^2 &=& \left({2+\D \ov 3}\right)^2 \ell (\ell+2)
+ 2 \left[ {\D(2+\D) \ov 3}  + \hgamma^2 (n_2-n_3)^2 \right] \nonumber\\
&\pm& 2 \sqrt{ \D^2 \left({2+\D \ov 3}\right)^2 \ell (\ell+2) +
\left[ {\D(2+\D) \ov 3} + \hgamma^2 (n_2-n_3)^2 \right]^2}\ .
\ea
Obviously the $\omega_{+,\ell n_2 n_3}^2$ are positive-definite
while, for $\hsigma=0$ ($\Delta=1$), the $\omega_{-,\ell n_2
n_3}^2$ are positive-semidefinite, with a zero mode occurring for
$\ell=0$. Therefore, the only potential source of instabilities
for these fluctuations is one of the $\omega_{-,\ell n_2 n_3}^2$
becoming negative for some value of $\hsigma$. This would only be
possible if the inequality
\be
\label{5-34} P(x) \equiv [\ell(\ell+2) - 24] x^2 + 4 [\ell(\ell+2)
+ 6] x + 4 [\ell(\ell+2) + 9 \hgamma^2 (n_2-n_3)^2] < 0\ ,
\ee
could be satisfied for some $x$ with $0 \leqslant x \leqslant 1$.
It is easily seen that this cannot happen for any values of
$\ell$. Turning to the $(\d \vec{r}_3,\d x)$ fluctuations, we find
that they satisfy the Schr\"odinger equations
\be
\left[ - {d^2 \ov d u^2} + F_r(u) - \left({2+\D \ov 3}\right)^2
\ell (\ell+2) - 4 \hgamma^2 (n_2-n_3)^2 \right] \d \vec{r}_{3,\ell
n_2 n_3} (u) = 0 \ ,
\label{dg1}
\ee
and
\be
\left[ - {d^2 \ov d u^2} + F_x(u) - \left({2+\D \ov 3}\right)^2
\ell (\ell+2) - 4 \hgamma^2 (n_2-n_3)^2 \right] \d x_{\ell n_2
n_3} (u) = 0 \ .
\label{dg2}
\ee
To examine them, we may consider the following cases.
\begin{itemize}
\item Zero rotation ($r_0=0$). In the absence of rotation
parameters (in which case $F_r=F_x=-1$), the $\d \vec{r}_3$ and
$\d x$ fluctuations obey the same Schr\"odinger equation with a
$u$--independent potential. Introducing an $e^{-{\rm i} \omega u}$
dependence, we easily obtain the spectrum
\be
\omega^2_{r, \ell n_2 n_3} = 1 + 4 \hgamma^2 (n_2-n_3)^2 +
\left({2+\D \ov 3}\right)^2 \ell (\ell+2)\ ,
\ee
whence we verify that the $\omega^2_{r, \ell n_2 n_3}$ are
manifestly positive-definite, signifying stability against small
perturbations in these directions. The $\omega^2_{r,\ell n_2 n_3}$
are increasing functions of $\hgamma$ but decreasing functions of
$\hsigma$ with the last term ranging from $\ell (\ell+2)$ at
$\hsigma=0$ to ${4 \ov 9}\ell (\ell+2)$ at $\hsigma = {1 \ov 2
\sqrt{3}}$. Note also that no zero mode is possible.

\item Non-zero rotation parameters ($r_0 \ne 0$). In this case,
the Schr\"odinger equations \eqn{dg1} and \eqn{dg2} have periodic
potentials $F_r$ and $F_x$, respectively, given by \eqn{4-44}
(with the parameter $a$ in \eqn{4-38}) and fixed eigenvalue
depending on the deformation parameters $\hat \gamma$ and $\hat
\s$ as well as on the quantum numbers $\ell$ and
$n_{2,3}$.\footnote{Exactly the same equations appeared in
\cite{bs-pp} in the light-cone quantization of strings moving in
these PP-wave backgrounds (for $\hat \g=\hat \s=0$).} From the
shape of the potentials and since the fixed eigenvalue is
non-negative, we infer stability. However, since the potentials
are periodic the spectrum is continuous with mass gaps, which tend
to zero as the parameter $a\to 0$. Hence, for fixed quantum
numbers $\ell$ and $n_{2,3}$, there exist ranges of the
deformation parameters $\hat\g$ and $\hat\s$ for which any other
solution than the vanishing one is not allowed.

\end{itemize}

\no In conclusion, despite the fact that $\s$--deformations render
the giant graviton states energetically unfavorable, the
small-fluctuation analysis indicates that these objects are
perturbatively stable in the range of $\hsigma$ where they are
allowed to exist in the first place. Presumably, the effect of
$\s$--deformations renders the giant gravitons metastable rather
than unstable. To further investigate this aspect, one may seek
``bounce-like'' instanton solutions connecting the giant and the
point graviton and calculate the tunneling probability by standard
WKB methods. In doing so, one must appropriately take into account
possible fermionic zero modes resulting from the breaking of
supersymmetries by the instanton, which tend to suppress the
tunneling rate.

\subsubsection{Dual giant gravitons}

We next consider dual giant gravitons which, as we recall, are
independent of the deformation. Now, the perturbation can be
described by keeping the gauge choice as in \eqn{5-a1} and
perturbing the embedding according to
\be
\label{5-a7} v = - \n u + \delta
v(u,\tilde{\psi},\tilde{\phi}_2,\tilde{\phi}_3)\ ,\quad
\tilde{\rho} = \tilde{\rho}_0 + \delta
\tilde{\rho}(u,\tilde{\psi},\tilde{\phi}_2,\tilde{\phi}_3)\ ,\quad
\vec{y}_4 =
\d\vec{y}_4(u,\tilde{\psi},\tilde{\phi}_2,\tilde{\phi}_3)\ .
\ee
Repeating the same steps as before, we find that the quadratic
term in the action of the fluctuations reads
\ba
\label{5-a8} S_2 &=& - {M \tilde{\rho}_0^2 \ov 4 \pi^2} \int du
d\tilde{\Omega}_3 \biggl[ (1+4|\hbeta|^2)\d\vec{y}_4^2 - {4 \ov
\tilde{\rho}_0} \d \tilde{\rho} \partial_u \d v \nonumber\\
&& \qq\qq\qq\qq + {1 \ov \tilde{\rho}_0^2} \d v ( \partial_u^2 -
\D_{{\rm S}^3} ) \d v + \d \vec{y}_5 \cdot ( \partial_u^2 -
\D_{{\rm S}^3} ) \d \vec{y}_5 \biggr]\ ,
\ea
where we introduced the shorthand $\d \vec{y}_5 = (\d
\vec{y}_4,\d\tilde{\rho})$. We see that the deformation enters
only through a modification of the mass term of $\d\vec{y}_4$.
Introducing an $e^{-{\rm i} \omega u}$ time dependence and
expanding the fluctuations on ${\rm S}^3$ as before, we find the
fluctuation spectrum
\be
\label{5-a9} \omega_{\pm,\ell n_2 n_3}^2 = \ell (\ell+2) + 2 \pm 2
\sqrt{
 \ell (\ell+2) + 1}\ ,
\ee
and
\be
\label{5-a10} \omega^2_{y, \ell n_2 n_3} = 1 + 4 |\hbeta|^2 + \ell
(\ell+2)\ ,
\ee
for the $(\d v,\d \tilde{\rho})$ and $\d\vec{y}_4$ fluctuations
respectively. This spectrum is manifestly positive-semidefinite,
with $\omega_{-,\ell n_2 n_3}^2$ having the expected zero mode for
$\ell=0$. We conclude that the deformation does not affect the
stability of dual giant gravitons, its sole effect being just a
raise of the energy of the $\d\vec{y}_4$ fluctuations.

\section{Probing the deformed geometry with Wilson loops}
\label{sec6}

To further investigate the effects of $\b$--deformations we now
turn to another, completely different, direction, of more
phenomenological nature. Namely, we consider the potential for a
static heavy $q\bar{q}$ pair in the dual gauge theory, extracted
from the expectation value of a rectangular Wilson loop extending
along the Euclidean time direction and one space direction. On the
gravity side, the Wilson loop expectation value is calculated
\cite{maldaloop,wilsonloopTemp} by minimizing the Nambu--Goto
action for a fundamental string propagating into the dual
supergravity background, whose endpoints are constrained to lie on
the two sides of the Wilson loop. Below, we first briefly review
the procedure for calculating Wilson loops in general supergravity
backgrounds of interest \cite{bs,hsz} and then we apply it to the
case of $\s$--deformations of the Coulomb branch.

\subsection{General formalism}

As stated above, the calculation of a Wilson loop in the gravity
approach amounts to extremizing the Nambu--Goto action (taking
into account the contribution of the NSNS 2-form if necessary) for
a string propagating in the dual geometry whose endpoints trace
the loop. To describe the propagation of the string, we first fix
reparametrization invariance by taking $(\tau,\s) = (t,x)$. We
next need to find a suitable ansatz that is sensitive to
$\b$--deformations. Specializing to a radial trajectory, it is
easily seen that the only available choice for the embedding
is\footnote{If, for instance, $\th=\pi/2$, then the resulting
Wilson loop potentials are identical to those for the $\cN=4$
undeformed theory computed in \cite{bs}. Also, to conform with
standard notation in the literature, we use $u$ instead of $r$ in
the Wilson-loop computations.}
\ba
\label{6-1} r=u(x)\ ,\quad \th=0\ ,\quad \psi={\pi\ov 4}\ ,\quad
\phi_1=\const\ , \quad \phi_2=\phi_3=\const\ ,\quad  {\rm rest} =
\const\ .
\ea
We next pass to the Euclidean using the analytic continuation $t
\to {\rm i} t$.\footnote{In extending this to the
finite-temperature case, one also has to take $r_0 \to -{\rm i}
r_0$ due to the presence of nonzero metric components $G_{ti} \sim
r_0 dt d\phi_i$.} Then, the Nambu--Goto action is found to be
\be
\label{6-2}
S = \frac {T}{2\pi} \int dx \sqrt{ f(u)/R^4 + g(u) u^{\prime 2} } \ ,
\ee
where the prime denotes a derivative with respect to $x$ and
\be
\label{6-3}
f(u) = R^4 G_{tt} G_{xx}\ ,\qq g(u) = G_{tt} G_{uu}\ .
\ee
We do not need to consider at all the contribution of the NSNS
2-form, since it is vanishing due to the ansatz \eqn{6-1} and to the
fact that, for the extremal D3-brane distributions considered here,
the 2-forms $C^i_3$ generating terms for the NSNS 2-form through
\eqn{2-9} vanish identically.

\no Since the action \eqn{6-2} does not explicitly depend on $x$,
it leads to the first integral $u_0$, identified with the turning
point of the solution. Solving the corresponding first-order
equation for $x$ in terms of $u$ we find that the linear
separation of the quark and antiquark is
\be
\label{6-4}
L= 2 R^2 f^{1/2}(u_0) \int^\infty_{u_0} du \sqrt{g(u)\ov f(u) [f(u) - f(u_0)]}\ .
\ee
The energy of the configuration is given by the action \eqn{6-2} divided by $T$.
Subtracting the self-energy contribution, we obtain
\be
\label{6-7}
E = {1 \ov \pi} \int^\infty_{u_0} du \left[\sqrt{g(u) f(u)\ov f(u) - f(u_0)} - \sqrt{g(u)}\right]
- {1\ov \pi} \int^{u_0}_{u_{\rm min}} du \sqrt{g(u)}\ ,
\ee
where $u_{\rm min}$ is the minimum value of $u$ allowed by the
geometry. In specific examples, we are supposed to solve for the
auxiliary parameter $u_0$ in terms of the separation distance
$L$. Since this cannot be done explicitly except for some special
cases, in practice one regards Eq. \eqn{6-4} as a parametric equation
for $L$ in terms of the integration constant $u_0$. Combining it
with Eq. \eqn{6-7} for $E$, one can then determine the behavior
of the potential energy of the configuration in terms of the
quark-antiquark separation.

\subsection{Application: $\s$--deformations of the Coulomb branch}

As an application of the above, we extend the results of \cite{bs}
for the undeformed theory, by considering the behavior of the static
$q\bar{q}$ potential for the case of pure $\s$--deformations
($\hgamma=0$) of the Coulomb branch of the gauge theory at zero
temperature. This was previously examined in \cite{ahn2}, where most
of the qualitative features of the potentials were extracted
numerically. Here, we pursue a more careful analysis, which allows
us to discover certain features previously unnoticed.

\no For the analysis that follows, it is convenient to use the
single dimensionful parameter $r_0$ of the theory to switch to
dimensionless variables. To do so, we set
\be
u \to r_0 u\ ,\qq u_0 \to r_0 u_0\ ,
\ee
and
\be
\label{6-13} L \to {R^2 \ov r_0} L\ , \qq E \to {r_0 \ov \pi} E\ .
\ee
Also, to keep the discussion at a reasonable length, we restrict to
the cases of two equal rotation parameters and one rotation
parameter. The results are presented below.

\subsubsection*{Two equal rotation parameters (sphere)}

For the case of two equal rotation parameters, the effective metric
for the trajectories \eqn{6-1} reads
\be ds^2 = \cH^{1/2} H^{-1/2} (
-  dt^2 + d x^2 ) + \cH^{1/2} H^{1/2} du^2\ ,
\ee
with
\be
\cH = 1 + {\hsigma^2 u^2 \ov u^2 - r_0^2}\ , \qq H = {R^4 \ov u^2 (
u^2 - r_0^2 )}\ .
\ee
Performing the analytic continuation, calculating the functions $f$
and $g$ according to \eqn{6-3}, inserting into \eqn{6-4} and
\eqn{6-7} and switching to dimensionless variables, we find the
following exact expressions for $L$ and $E$ in terms of complete
elliptic integrals\footnote{In the following we adopt the notation and use
properties of elliptic integrals as in \cite{tipologio} and \cite{BF}.}
\ba
\label{6-17}
L &=& 2 u_0 \sqrt{u_0^2 - {1 \ov 1+\hsigma^2}}
\int_{u_0}^\infty {du \ov u \sqrt{(u^2 - u_0^2) ( u^2 - 1) ( u^2 + u_0^2 - {1 \ov 1+\hsigma^2} ) } } \nonumber\\
&=& { 2 u_0 \ov \sqrt{(u_0^2-{1 \ov 1+\hsigma^2})(2 u_0^2-{1 \ov
1+\hsigma^2})} } \left[ \elPi(a^2,k) - \elK(k) \right]\ ,
\ea
and
\ba
\label{6-18}
\!\!\!\!\!\!\!\!E &=& \sqrt{1 + \hsigma^2} \int_{u_0}^\infty du
\sqrt{{u^2 - {1 \ov 1+\hsigma^2} \ov u^2 - 1}}
\left( \sqrt{ { u^2 ( u^2 - {1 \ov 1+\hsigma^2} )
\ov (u^2 - u_0^2)( u^2 + u_0^2 - {1 \ov 1+\hsigma^2} ) } } - 1 \right) \nonumber\\
&& -\:\: \sqrt{1 + \hsigma^2}
\int_{1}^{u_0} du \sqrt{{ u^2 - {1 \ov 1+\hsigma^2} \ov u^2 - 1}} \nonumber\\
&=& \sqrt{1+\hsigma^2} \left\{  \sqrt{2 u_0^2-{1 \ov 1+\hsigma^2}}
\left[ a^2 \elK(k) - \elE(k) \right] + \elE(c)-{\hsigma^2 \ov
1+\hsigma^2}\elK(c) \right\}\ ,
\ea
where
\be
k^2 = {u_0^2 + {\hsigma^2 \ov 1+\hsigma^2} \ov 2 u_0^2 - {1 \ov
1+\hsigma^2}}\ ,\qq a^2 = {u_0^2 - {1 \ov 1+\hsigma^2} \ov 2 u_0^2 -
{1 \ov 1+\hsigma^2}}\ , \qq c={1 \ov \sqrt{1 + \hsigma ^2}}\ .
\ee
\begin{figure}[!t]
\begin{center}
\begin{tabular}{ccc}
\includegraphics [height=3.6cm]{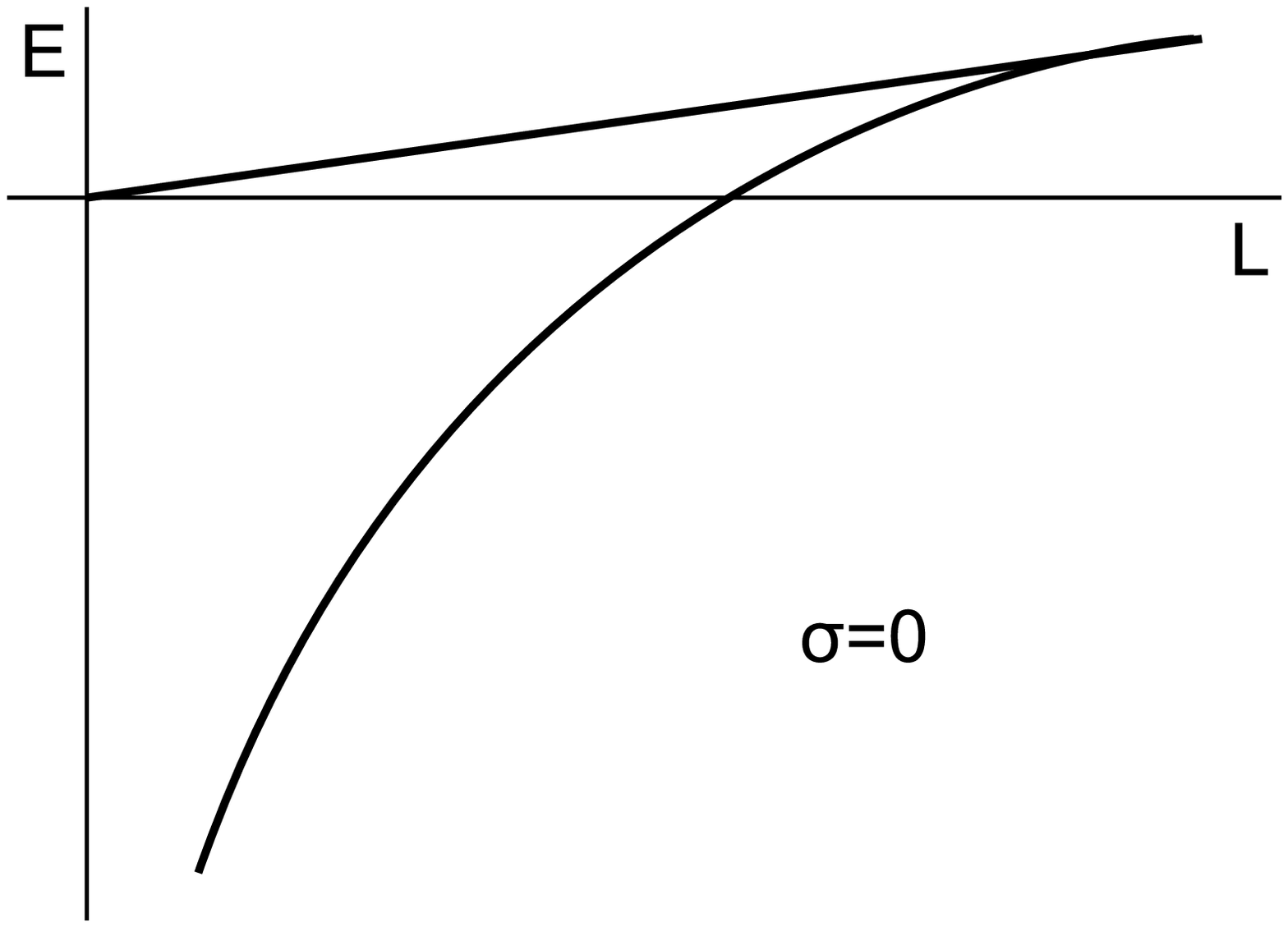}
& \includegraphics [height=3.77cm]{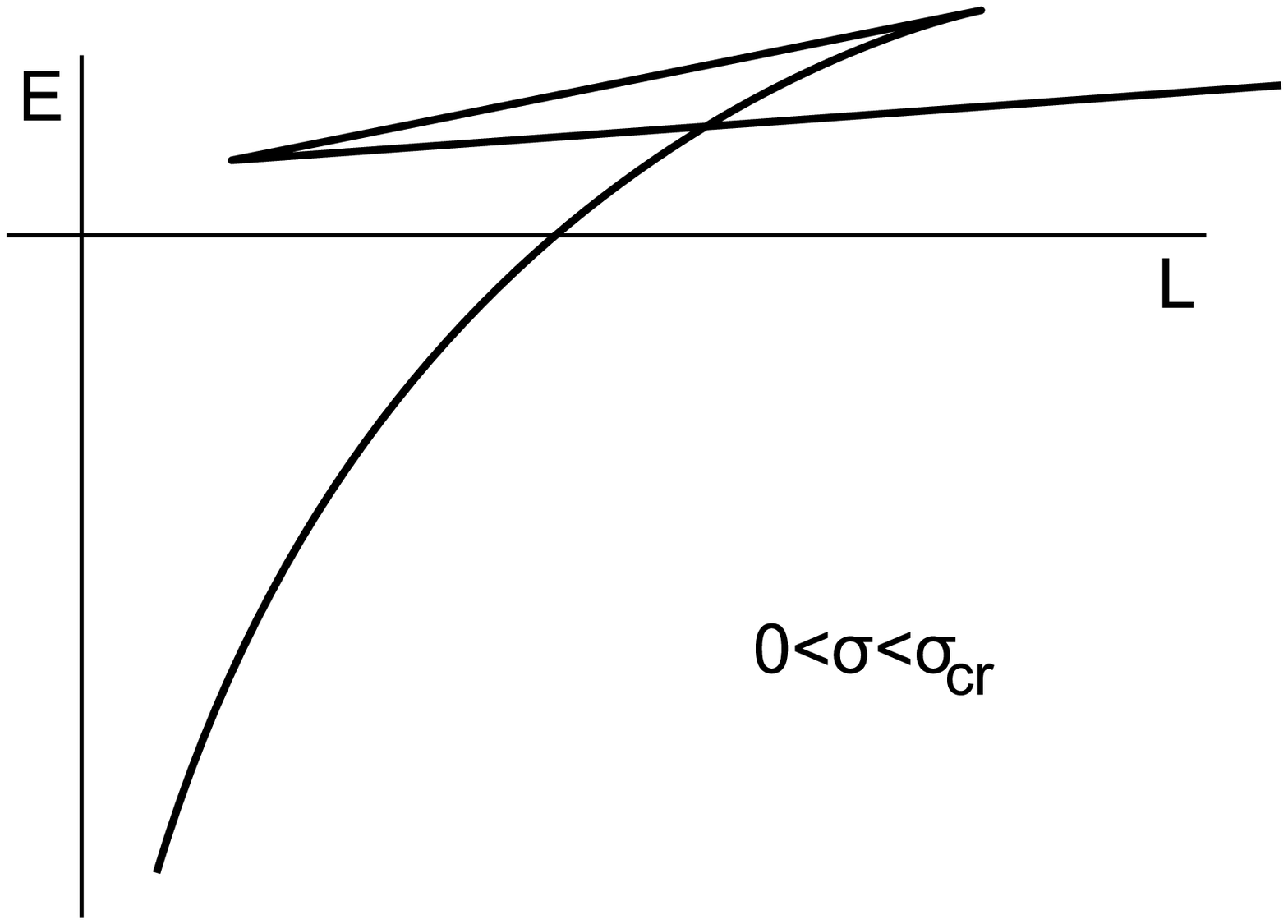}
& \includegraphics [height=4.15cm]{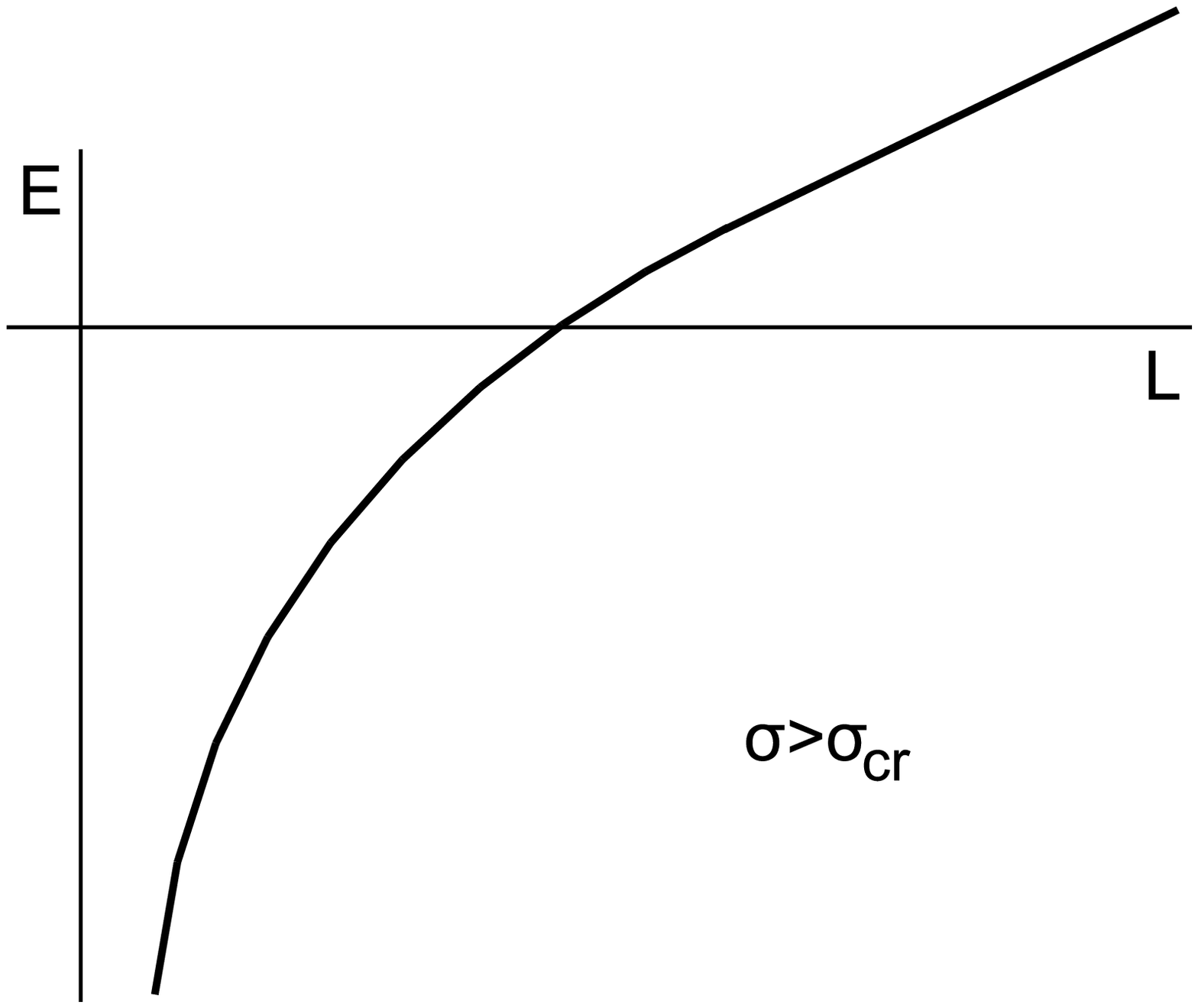} \\
(a) & (b) & (c)
\end{tabular}
\end{center}
\caption{Energy as a function of length for the case of two equal
rotation parameters.}
\end{figure}The resulting plots of $E$ versus $L$ are shown in Fig. 2.
For $\hsigma=0$, the behavior is that for the undeformed case
\cite{bs}. As $\hsigma$ is turned on, the length and energy curves
closely resemble the van der Waals isotherms for a statistical
system with $u_0$, $L$ and $E$ corresponding to volume, pressure
and Gibbs potential respectively (see, for instance,
\cite{Callen}). In the region below a critical point, $\hsigma <
\hsigma_{\rm cr}$, the behavior is analogous to that of the
statistical system at $T<T_{\rm cr}$. Namely, the potential energy
$E$ (i) starts out Coulombic at small distances, (ii) becomes a
triple-valued function of $L$ with the state of lowest energy
corresponding to the initial branch, (iii) passes a
self-intersection point after which it becomes a triple-valued
function of $L$ with the state of lowest energy corresponding to
the second branch and (iv) returns to being a single-valued
function of $L$ with approximately linear behavior. By standard
arguments, the physical path in the length and energy curves must
correspond to the physical isotherms of the statistical system,
with the self-intersection point in the energy curve indicating a
first-order phase transition with order parameter $u_0$. In the
region above the critical point, $\hsigma > \hsigma_{\rm cr}$, the
behavior is analogous to that of the statistical system at
$T>T_{\rm cr}$. Now the energy is single-valued throughout and the
first-order phase transition has degenerated into a second-order
one between a Coulombic phase and a confining phase with a linear
potential, as we will show below. The above critical behavior is
similar to that found in \cite{bs} for a different system, namely
for the undeformed theory at finite temperature and non-zero
chemical potential. Since the two cases are in complete analogy,
we refer the reader to that work for several related computational details.

\no
To determine the critical value of $\hsigma$, we consider the
derivative of the separation length $L$ with respect to $u_0$.
This derivative is proportional to the function
\be
\label{6-20} f(u_0;\hsigma)={1\ov 1+2 \hsigma^2}\left\{ \hsigma^2 +
2 (1+\hsigma^2) u_0^2  - (1+\hsigma^2)^2 u_0^2 [ 2 u_0^2 \elK (k) +
(1-u_0^2) \elE(k) ] \right\}\ .
\ee
This function has a single zero corresponding to a global maximum
of the length for $\hsigma=0$, two zeros corresponding to a local
minimum and a local maximum of the length for
$0<\hsigma<\hsigma_{\rm cr}$, and no extrema for $\hsigma >
\hsigma_{\rm cr}$. To calculate $\hsigma_{\rm cr}$, we proceed by
expanding $f(u_0;\hsigma)$ around $u_0=1$ corresponding to the
modulus $k=1$; although this procedure is not \emph{a priori}
valid, it will be justified by our final result. We have
\be
\label{6-21} f = - \hsigma^2 -4 x \left[ 8+15\hsigma^2 +
(2+\hsigma^2) \ln x \right] + {\cal O}(x^2)\ , \qq x \equiv
{1+\hsigma^2\ov 8(1+2 \hsigma^2)}\ (u_0 - 1)\ ,
\ee
and
\be
\label{6-22}
{\partial f \ov \partial x} = -4 \left[ 10+ 16 \hsigma^2 +  (2+\hsigma^2) \ln x \right] + {\cal O}(x) \ .
\ee
Setting $f=0$ gives the transcendental equation
\ba
-a \ln a = {\hsigma^2\ov 4(2+\hsigma^2)}\
 e^{ 8+15\hsigma^2\ov 2+\hsigma^2} \leqslant e^{-1} \ ,\qq
a = x e^{ 8+15\hsigma^2\ov 2+\hsigma^2} \ .
\label{6-27}
\ea
This equation has two solutions $a_1,a_2$ if $\hsigma <
\hsigma_{\rm cr}$, one solution $a_1=a_2=a_c$ if $\hsigma =
\hsigma_{\rm cr}$ and no solutions if $\hsigma > \hsigma_{\rm
cr}$. The critical value $\hsigma_c$ is obtained when the above
inequality is saturated in which case the solution is $a=e^{-1}$.
It turns out that in this case we have in addition that $\partial
f / \partial x = 0$. The corresponding transcendental equation is
\be
\label{6-24} z e^{z+5} = 44\ ,\qq z \equiv 11 {\hsigma_{\rm
cr}^2\ov 2+\hsigma_{\rm cr}^2}\ .
\ee
This gives $z\simeq 0.235$, whence $\hsigma_{\rm cr}\simeq 0.209$, which
clearly is small enough to validate in retrospect the approximation method we
used to compute it.
Having a solution to \eqn{6-27} we obtain from the definition in
\eqn{6-21} that the corresponding critical value(s) for $u_0$ are
given by
\be
u_{0i}=1+{a_i\ov 8} {1+2 \hsigma^2\ov 1+\hsigma^2} e^{-{
8+15\hsigma^2\ov 2+\hsigma^2}} \ ,\qq i=1,2\ ,
\ee
for $\hsigma \leqslant \hsigma_{\rm cr}$.

\no
Finally, note that for $u_0 \to 1$ (large $L$) we recover the usual Coulombic
behavior, while for $u_0 \to 1$ we find the asymptotics
\be
\label{conf} L \simeq {\hsigma \ov \sqrt{1+2\hsigma^2}} \ln {1 \ov
u_0-1}\ ,\qq E \simeq {\hsigma^2 \ov 2 \sqrt{1+2\hsigma^2}} \ln {1
\ov u_0-1}\ .
\ee
Combining these expressions, we obtain the potential
\be
E \simeq {\hsigma \ov 2} L\ ,
\ee
which demonstrates the linear confining behavior claimed earlier
on, as long as $\hat\s >0$. This linear potential was also found
in the studies of \cite{ahn2} where, however, the critical
behavior found here was missed. Using the first of \eqn{conf} and
reinstating dimensional units according to the first \eqn{6-13},
we find that confinement sets in at length scales $L \gtrsim
{\hsigma \ov \sqrt{1+2\hsigma^2}} {R^2 \ov r_0}$.

\subsubsection*{One rotation parameter (disc)}

For the case of one rotation parameter, the effective metric for the trajectories
\eqn{6-1} reads
\be
ds^2 = \cH^{1/2} H^{-1/2} ( -  dt^2 + d x^2 ) + \cH^{1/2} H^{1/2}
du^2\ ,
\ee
with
\be
\cH = 1 + {\hsigma^2 u^2 \ov u^2 + r_0^2}\ ,\qq H = {R^4 \ov u^2 (
u^2 + r_0^2 )}\ .
\ee
Proceeding in the same way as before, we find that the length and
potential energy are given by
\ba
\label{6-14}
L &=& 2 u_0 \sqrt{ u_0^2 + {1 \ov 1+\hsigma^2} } \int_{u_0}^\infty  {du \ov u \sqrt{ (u^2 - u_0^2)(u^2+1)(u^2 + u_0^2 + {1 \ov 1+\hsigma^2})} }\nonumber\\
&=& { 2 u_0 \ov \sqrt{(u_0^2+{1 \ov 1+\hsigma^2})(2 u_0^2+{1 \ov
1+\hsigma^2})} } \left[ \elPi(a^2,k) - \elK(k) \right]\ ,
\ea
and
\ba
\label{6-15}
\!\!\!\!\!\!\!\!\!\!\!\!\!\!\!\!\!\!\!\! E &=& \sqrt{1 + \hsigma^2}
\int_{u_0}^\infty du \sqrt{ { u^2 + {1 \ov 1+\hsigma^2} \ov u^2 + 1} }
\left( \sqrt{ {u^2 (u^2 + {1 \ov 1+\hsigma^2})
\ov (u^2 - u_0^2) ( u^2 + u_0^2 + {1 \ov 1+\hsigma^2}) } } - 1 \right) \nonumber\\
&& - \:\: \sqrt{1 + \hsigma^2} \int_0^{u_0} du \sqrt{ { u^2 + {1 \ov 1+\hsigma^2} \ov u^2 + 1} }
\nonumber\\
&=& \sqrt{1+\hsigma^2} \left\{  \sqrt{2 u_0^2+{1 \ov 1+\hsigma^2}}
\left[ a^2 \elK(k) - \elE(k) \right] + \elE(c)-{1 \ov
1+\hsigma^2}\elK(c) \right\}\ .
\ea
where
\be k^2 = {u_0^2 - {\hsigma^2 \ov 1+\hsigma^2} \ov 2
u_0^2 + {1 \ov 1+\hsigma^2}}\ ,\qq a^2 = {u_0^2 + {1 \ov
1+\hsigma^2} \ov 2 u_0^2 + {1 \ov 1+\hsigma^2}}\, \qq c={\hsigma \ov
\sqrt{1 + \hsigma ^2}}\ .
\ee
\begin{figure}[!t]
\begin{center}
\begin{tabular}{c}
\includegraphics [height=6cm]{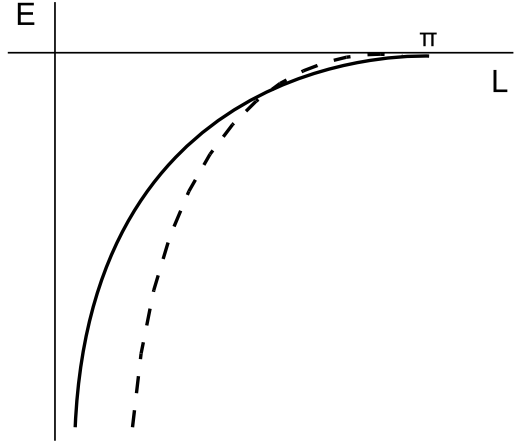}
\end{tabular}
\end{center}
\caption{Energy as a function of length for the case of one
rotation parameter, shown for $\hsigma=0$ (solid) and $\hsigma>0$
(dashed).}
\end{figure}Fig. 3 shows the resulting plots of $E$ versus $L$. For $u_0 \gg 1$ (small $L$) we recover the
standard Coulombic behavior \cite{maldaloop} enhanced by the factor $\sqrt{1+\hat \s^2}$,
while for $u_0 \to 0$ we find the asymptotics
\be
L \simeq \pi - 2 \elE( {\rm i} \hsigma) u_0\ ,\qq E \simeq - {1
\ov 2} \elE( {\rm i} \hsigma) u_0^2\ , \label{gsle}
\ee
which reduce to those in \cite{bs} for $\hsigma = 0$. Combining
these expressions, we obtain
\be
E \simeq - {(\pi - L)^2 \ov 8 \elE({\rm i} \hsigma)}\ ,
\ee
which shows that there is complete screening at the screening
length $L_{\rm c} = \pi$ that is invariant under
$\s$--deformations and the same as the one found \cite{bs} for the
undeformed case. The effect of $\s$--deformations is to enhance
the quark-antiquark force for small separations and to suppress it
for separations close to the screening length.\footnote{Note the
crossover behavior for the curves with $\hat \s=0$ and $\hat \s>0$
in our Fig. 3 at a certain length. This is understood by our
analytic result \eqn{gsle} as well as the enhanced Coulombic
behavior in the UV noted above. In that respect we disagree with
the shape of the potential presented in Fig. 3 of \cite{ahn2}.}
All of our results reduce smoothly to those in \cite{bs} for
$\hat\s=0$.

\no We finally remark that it would be very interesting to examine
the stability of the string trajectories used for calculating the
quark--antiquark potential in the deformed theories. For instance,
in the undeformed theory with two rotation parameters, the potential
for our trajectory \eqn{6-1} with $\th=0$ (shown in Fig. 2(a)) is a
double-valued function while the potential for a trajectory with
$\th=\pi/2$ (which is insensitive to deformations and thus has not
been considered here) exhibits a confining behavior at large
distances \cite{bs}. It was found in \cite{ASSiampos} that the upper
branch in the $\th=0$ case as well as the region giving rise to
linear behavior in the $\th=\pi/2$ case are actually unstable under
small fluctuations, the latter fact being in accordance with our
physical expectation about the absence of confinement in $\cN=4$
SYM. In the presence of deformation, the potential for $\th=0$ gives
the behavior shown in Fig. 2(b,c), while the potential for
$\th=\pi/2$ stays invariant. It is then important to ask whether the
confining regions of these potentials are stable, as the $\cN=1$
supersymmetry of the $\s$--deformed theories actually leads us to
expect a confining behavior at large distances, at least in some
regions of the moduli space. We note that the question of stability
is meaningful even in the $\th=\pi/2$ case since, although the
classical string solution is independent of the deformation, small
fluctuations about it are {\em not}. We hope to report on work in
that direction in the future \cite{ASSi2}.

\section{Conclusions}
\label{sec7}

In this paper, we have explicitly applied the Lunin--Maldacena
construction of complex marginal deformations of supergravity
solutions to a class of general Type IIB backgrounds that include
the gravity duals of $\cN=4$ gauge theories at finite temperature
and R-charge chemical potentials and at the Coulomb branch. For
these theories, we have concentrated on three simple cases of the
general solution, we have presented in full detail the
marginally-deformed metrics, and we have checked that their
thermodynamics (for the rotating case) are the same as for the
undeformed ones, as they should.

\no Having constructed the marginally-deformed spacetimes, we
considered their Penrose limits for the multicenter case along a
certain class of geodesics inside the angular part of the
geometry. Besides recovering familiar results, we have extended
them to take account of the presence of $\s$--deformations, which
has not been considered in the literature up to date (with the
exception of the archetypal example of \cite{LM}), as well as for
the presence of rotation (previously examined in \cite{bs-pp} for
the undeformed solutions). We have also considered Penrose limits
along a non-BPS geodesic.

\no We next turned to a study of the giant gravitons supported on
the PP--wave spacetimes just constructed, using the simple example
of the PP--wave of \cite{LM} but taking the $\s$--deformation into
account. For that case we found that, unlike $\g$--deformations,
$\s$--deformations lift the degeneracy between the giant and the
point graviton but, nevertheless, the former remains stable under
small perturbations. We also showed that, for the particular
geodesic under consideration, this result is unaffected by the
presence of rotation. We also considered dual giant gravitons, in
which case the situation remains qualitatively unchanged by the
deformation. It would be interesting to generalize these studies
to more complicated PP--waves and to seek giant graviton solutions
in the full $\s$--deformed geometry.

\no Finally, we considered the standard Wilson-loop calculation of
interquark potentials and screening lengths for the dual gauge
theory. Here we have considered the static heavy-quark potential in
the case of the Coulomb branch of the $\s$--deformed gauge theory,
and for the two D3-brane distributions under consideration we found
a linear confining potential and a screened Coulombic potential
respectively. For the first case, we elucidated on the nature of the
transition to the confining phase and we calculated the critical
value of the deformation parameter, while for the second case we
demonstrated invariance of the screening length under the
deformation. As noted in the text, it is very interesting to examine
the stability of the string trajectories used to calculate
potentials in the deformed theory using the formalism developed in
\cite{ASSiampos}. Also it is of some interest to extend these
results to the theory at finite temperature and chemical potential.

\vskip 1cm

\centerline{ \bf Acknowledgments}

\no K.~S. acknowledges support provided through the European
Community's program ``Constituents, Fundamental Forces and
Symmetries of the Universe'' with contract MRTN-CT-2004-005104,
the INTAS contract 03-51-6346 ``Strings, branes and higher-spin
gauge fields'' and the Greek Ministry of Education programs $\rm \P
Y\Th A\G OPA\S$ with contract 89194.

\appendix

\section{T-- and S--duality rules}

Here, we state our conventions for the T-- and S--duality
transformations used in Section 2. Starting from T--duality, we
consider a Type II configuration characterized by the metric
$G_{MN}$, the NSNS 2-form $B_2$, the dilaton $\Phi$ and the RR
$p$-forms $A_p$ and we wish to T--dualize along an isometry
direction, say $y$. Splitting the coordinates as $x^M=(x^\mu,y)$,
we decompose the metric, the NSNS 2-form and the RR $p$--forms
into the quantities
\be
\label{a-1} a_1 \equiv G_{y \mu} dx^\mu\ ,\qq \phi \equiv G_{yy}\
,\qq b_2 \equiv \frac{1}{2} B_{\mu\nu} dx^\mu \wedge dx^\nu\ ,\qq
b_1 \equiv B_{y \mu} dx^\mu\ ,
\ee
and
\be
\label{a-2}
\a_p \equiv {1 \ov p!} A_{\mu \ldots \nu \rho} dx^\mu \wedge \ldots
\wedge dx^\nu \wedge dx^\rho\ ,\qq \b_{p-1} \equiv {1 \ov (p-1)!}
A_{\mu \ldots \nu y} dx^\mu \wedge \ldots \wedge dx^\nu\ ,
\ee
so that, in particular, $B_2 = b_2 - b_1 \wedge dy$ and $A_p =
\a_p + \b_{p-1} \wedge dy$. Under T--duality, the NSNS fields
transform among themselves by the usual Buscher rules
\cite{buscher} while the RR fields transform by terms involving
both NSNS and RR fields \cite{bho}. In our present notation, these
transformations can be written in the compact form
\ba
\label{a-3}
\hat{G}_{MN} dx^M dx^N &=& G_{\mu\nu} dx^\mu dx^\nu + \phi^{-1}
\left[ (dy + b_1)^2 - a_1^2 \right]\, \nonumber\\
\hat{B}_2 &=& b_2 - \phi^{-1} a_1 \wedge (dy + b_1)\ ,\nonumber \\
e^{2\hat{\Phi}} &=& \phi^{-1} e^{2\Phi}\ , \\
\hat{A}_p &=& \b_p + ( \a_{p-1} - \phi^{-1} \b_{p-2} \wedge a_1 ) \wedge ( dy + b_1)\ .
\nonumber
\ea
This form of the T--duality rules is particularly useful in the
computations of section \ref{sec2}, as it allows us to perform the
transformations directly in form notation.

\no We next consider an S--duality transformation of a Type IIB
configuration, which acts on the axion-dilaton $\tau = A_0 +
\rm{i} e^{-\Phi}$ and the NSNS and RR 2-forms as follows
\ba
\hat \tau = {a\tau+b\ov c \tau + d}\ ,\qq \left(
\begin{array}{c}
  -\hat B_2 \\
  \hat A_2 \\
\end{array}
\right) = (\L^T)\inv \left(
\begin{array}{c}
  -B_2 \\
  A_2 \\
\end{array}%
\right)\ ,\quad \L = \left(%
\begin{array}{cc}
  a & b \\
  c & d \\
\end{array}
\right)\ ,
\ea
leaving the Einstein-frame metric and the RR 5-form field strength
invariant. For the purpose of $\s$--deformations, we consider the
particular $SL(2,\mathbb{R})$ element
\be
\L = \left(
\begin{array}{cc}
  1 & 0 \\
  -\tilde \s & 1 \\
\end{array}%
\right)\ .
\ee
which, in addition, leaves the RR 2-form potential $A_2$
invariant. The resulting transformations of all fields, including
the string-frame metric and the RR 4-form, are written explicitly
as
\ba
\label{a-5}
\hat{G}_{MN} &=& \l^{1/2} G_{MN}\ , \nonumber\\
\hat{B}_2 &=& B_2 - \tilde{\s} A_2\ , \nonumber\\
e^{2\hat{\Phi}} &=& \l^2 e^{2 \Phi}\ , \\
\hat{A}_0 &=& \l^{-1} \left[ A_0 (1 - \tilde{\s} A_0) - \tilde{\s} e^{-2 \Phi} \right]\ , \nonumber\\
\hat{A}_4 &=& A_4 - {1 \ov 2} \tilde{\s} A_2 \wedge A_2\ . \nonumber
\ea
where
\be
\label{a-4} \l \equiv (1 - \tilde{\s} A_0)^2 + \tilde{\s}^2 e^{-2
\Phi}\ .
\ee



\begin{thebibliography}{99}

\renewcommand{\baselinestretch}{1}
\normalsize

\bibitem{adscft} J.M.~Maldacena,
Adv.\ Theor.\ Math.\ Phys.\ {\bf 2} (1998) 231, Int.\ J.\ Theor.\ Phys.\  {\bf 38} (1999) 1113,
{\tt hep-th/9711200}.\hfill\break
S.S.~Gubser, I.R.~Klebanov and A.M.~Polyakov,
Phys.\ Lett.\ {\bf B428} (1998) 105, {\tt hep-th/9802109}.\hfill\break
E.~Witten,
Adv.\ Theor.\ Math.\ Phys.\ {\bf 2} (1998) 253, {\tt hep-th/9802150}
and
Adv.\ Theor.\ Math.\ Phys.\  {\bf 2} (1998) 505,
{\tt hep-th/9803131}.

\bibitem{LS} R.G.~Leigh and M.J.~Strassler,
Nucl. Phys. {\bf B447} (1995) 95, {\tt hep-th/9503121}.

\bibitem{LM} O.~Lunin and J.~Maldacena,
JHEP {\bf 0505} (2005) 033, {\tt hep-th/0502086}.

\bibitem{frolov}
S.~Frolov,
JHEP {\bf 0505} (2005) 069, {\tt hep-th/0503201}.

\bibitem{frt}
S.A.~Frolov, R.~Roiban and A.A.~Tseytlin,
Nucl.\ Phys.\ {\bf B731} (2005) 1,\hfill\break
{\tt hep-th/0507021}.


\bibitem{marginal-generalizations}
R.C.~Rashkov, K.S.~Viswanathan and Y.~Yang,
Phys.\ Rev.\ {\bf D72} (2005) 106008,
{\tt hep-th/0509058}.

\bibitem{marginal-otherbackgrounds}
C.h.~Ahn and J.F.~V\'azquez-Poritz,
JHEP {\bf 0507} (2005) 032,
{\tt hep-th/0505168}
and
Class.\ Quant.\ Grav.\ {\bf 23} (2006) 3619,
{\tt hep-th/0508075}.\hfill\break
J.P.~Gauntlett, S.~Lee, T.~Mateos and D.~Waldram,
JHEP {\bf 0508}, 030 (2005)
{\tt hep-th/0505207}.

\bibitem{marginal-various}
S.A.~Frolov, R.~Roiban and A.A.~Tseytlin,
JHEP {\bf 0507} (2005) 045,
{\tt hep-th/0503192}
and
Nucl.\ Phys.\ {\bf B731} (2005) 1,
{\tt hep-th/0507021}.\hfill\break
U.~G\"ursoy and C.~N\'u\~nez,
Nucl.\ Phys.\ {\bf B725} (2005) 45, {\tt hep-th/0505100}.\hfill\break
N.P.~Bobev, H.~Dimov and R.C.~Rashkov,
{\em Semiclassical strings in Lunin-Maldacena background}, {\tt hep-th/0506063}.\hfill\break
J.G.~Russo,
JHEP {\bf 0509} (2005) 031,
{\tt hep-th/0508125}.\hfill\break
H.Y.~Chen and S.~Prem Kumar,
JHEP {\bf 0603} (2006) 051,
{\tt hep-th/0511164}.\hfill\break
L.F.~Alday, G.~Arutyunov and S.~Frolov,
JHEP {\bf 0606} (2006) 018,
{\tt hep-th/0512253}.\hfill\break
U.~G\"ursoy,
JHEP {\bf 0605} (2006) 014, {\tt hep-th/0602215}.\hfill\break
C.S.~Chu and V.V.~Khoze,
JHEP {\bf 0607} (2006) 011,
{\tt hep-th/0603207}.\hfill\break
R.~Minasian, M.~Petrini and A.~Zaffaroni,
JHEP {\bf 0612} (2006) 055,
{\tt hep-th/0606257}.

\bibitem{freedman}
S.S.~Pal,
Phys.\ Rev.\  {\bf D72} (2005) 065006, {\tt
hep-th/0505257}.\hfill\break
D.Z.~Freedman and U.~G\"ursoy,
JHEP {\bf 0511} (2005) 042, {\tt hep-th/0506128}.

\bibitem{berenstein}
D.~Berenstein and R.G.~Leigh,
JHEP {\bf 0001} (2000) 038,
{\tt hep-th/0001055}.\hfill\break
D.~Berenstein, V.~Jejjala and R.G.~Leigh,
Nucl.\ Phys.\ {\bf B589} (2000) 196,\hfill\break
{\tt hep-th/0005087}
and
Phys.\ Lett.\ {\bf B493} (2000) 162,
{\tt hep-th/0006168}.

\bibitem{dorey}
N.~Dorey,
JHEP {\bf 0408} (2004) 043,
{\tt hep-th/0310117}.\hfill\break
N.~Dorey and T.J.~Hollowood,
JHEP {\bf 0506} (2005) 036,
{\tt hep-th/0411163}.

\bibitem{ZanonPenati}
  A.~Mauri, S.~Penati, A.~Santambrogio and D.~Zanon,
  JHEP {\bf 0511} (2005) 024,
\hfill\break {\tt hep-th/0507282}.\hfill\break
  F.~Elmetti, A.~Mauri, S.~Penati and A.~Santambrogio,
  JHEP {\bf 0701} (2007) 026,
\hfill\break {\tt hep-th/0606125}.

\bibitem{trivedi}
P.~Kraus, F.~Larsen and S.P.~Trivedi,
JHEP {\bf 9903} (1999) 003,
{\tt hep-th/9811120}.

\bibitem{hsz} R.~Hern\'andez, K.~Sfetsos and D.~Zoakos,
JHEP {\bf 0603} (2006) 069, {\tt hep-th/0510132} 
and Fortsch.\ Phys.\ {\bf 54} (2006) 407, {\tt hep-th/0512158}.

\bibitem{maldaloop} J.M.~Maldacena,
Phys. Rev. Lett. {\bf 80} (1998) 4859, {\tt hep-th/9803002}.\hfill\break
S.J.~Rey and J.T.~Yee,
Eur. Phys. J. {\bf C22} (2001) 379, {\tt hep-th/9803001}.

\bibitem{ahn2} C.~Ahn and J.F.~V\'azquez-Poritz,
JHEP {\bf 0606} (2006) 061,
{\tt hep-th/0603142}.

\bibitem{cy}
M.~Cvetic and D.~Youm,
Nucl. Phys. {\bf B477} (1996) 449,
{\tt hep-th/9605051}.

\bibitem{rs}
J.G.~Russo and K.~Sfetsos,
Adv.\ Theor.\ Math.\ Phys.\  {\bf 3} (1999) 131,
{\tt hep-th/9901056}.

\bibitem{penrose}
R. Penrose, {\em Any space-time has a plane-wave as a limit},
Differential Geometry and Relativity, Reidel, Dordrecht, 1976.

\bibitem{sfepp}
  K.~Sfetsos,
  Phys.\ Lett.\   {\bf B324} (1994) 335,
  {\tt hep-th/9311010}.\hfill\break
  K.~Sfetsos and A.A.~Tseytlin,
  Nucl.\ Phys. {\bf B427} (1994) 245, {\tt hep-th/9404063}.

\bibitem{gueven}
R.~G\"uven,
Phys.\ Lett. {\bf B482} (2000) 255,
{\tt hep-th/0005061}.

\bibitem{blaupp}
M.~Blau, J.~Figueroa-O'Farrill, C.~Hull and G.~Papadopoulos,
JHEP {\bf 0201} (2002) 047,
{\tt hep-th/0110242}.

\bibitem{bmn}
D.~Berenstein, J.M.~Maldacena and H.~Nastase,
JHEP {\bf 0204} (2002) 013,\hfill\break {\tt hep-th/0202021}.

\bibitem{sadri}
D.~Sadri and M.M.~Sheikh-Jabbari,
Rev.\ Mod.\ Phys.\ {\bf 76} (2004) 853,\hfill\break {\tt
hep-th/0310119}.

\bibitem{np}
V.~Niarchos and N.~Prezas,
JHEP {\bf 0306} (2003) 015, {\tt hep-th/0212111}.

\bibitem{pp-deformed}
R.~de Mello Koch, J.~Murugan, J.~Smolic and M.~Smolic,
JHEP {\bf 0508} (2005) 072, {\tt hep-th/0505227}.\hfill\break
T.~Mateos,
JHEP {\bf 0508} (2005) 026, {\tt hep-th/0505243}.

\bibitem{bs-pp}
A.~Brandhuber and K.~Sfetsos,
JHEP {\bf 0212} (2002) 050,
{\tt hep-th/0212056}.

\bibitem{blauloughlin}
M.~Blau and M.~O'Loughlin,
Nucl.\ Phys.\ {\bf B654} (2003) 135, {\tt
hep-th/0212135}.\hfill\break
M.~Blau, M.~O'Loughlin, G.~Papadopoulos and A.A.~Tseytlin,
Nucl.\ Phys.\  {\bf B673} (2003) 57, {\tt hep-th/0304198}.

\bibitem{mcgreevy}
J.~McGreevy, L.~Susskind and N.~Toumbas,
JHEP {\bf 0006} (2000) 008,\hfill\break
{\tt hep-th/0003075}.

\bibitem{myers}
R.C.~Myers,
JHEP {\bf 9912} (1999) 022,
{\tt hep-th/9910053}.

\bibitem{dual-giants}
M.T.~Grisaru, R.C.~Myers and O.~Tafjord,
JHEP {\bf 0008} (2000) 040, {\tt hep-th/0008015}.\hfill\break
A.~Hashimoto, S.~Hirano and N.~Itzhaki,
JHEP {\bf 0008} (2000) 051, {\tt hep-th/0008016}.

\bibitem{giant-more}
S.R.~Das, A.~Jevicki and S.D.~Mathur,
Phys.\ Rev.\ {\bf D63} (2001) 044001,\hfill\break
{\tt hep-th/0008088}.\hfill\break
S.R.~Das, S.P.~Trivedi and S.~Vaidya,
JHEP {\bf 0010} (2000) 037,
{\tt hep-th/0008203}.\hfill\break
A.~Mikhailov,
JHEP {\bf 0011} (2000) 027,
{\tt hep-th/0010206}.\hfill\break
R.C.~Myers and O.~Tafjord,
JHEP {\bf 0111} (2001) 009,
{\tt hep-th/0109127}.\hfill\break
I.~Bena and D.J.~Smith,
Phys.\ Rev.\ {\bf D71} (2005) 025005,
{\tt hep-th/0401173}.

\bibitem{giant-sym}
V.~Balasubramanian, M.~Berkooz, A.~Naqvi and M.J.~Strassler,
JHEP {\bf 0204} (2002) 034,
{\tt hep-th/0107119}.\hfill\break
S.~Corley, A.~Jevicki and S.~Ramgoolam,
Adv.\ Theor.\ Math.\ Phys.\ {\bf 5} (2002) 809,
{\tt hep-th/0111222}.\hfill\break
D.~Berenstein,
Nucl.\ Phys.\ {\bf B675} (2003) 179,
{\tt hep-th/0306090}.\hfill\break
R.~de Mello Koch and R.~Gwyn,
JHEP {\bf 0411} (2004) 081,
{\tt hep-th/0410236}.

\bibitem{squashed-giants}
S.~Prokushkin and M.M.~Sheikh-Jabbari,
JHEP {\bf 0407} 077 (2004),
{\tt hep-th/0406053}.

\bibitem{unstable-giants}
R.~de Mello Koch, N.~Ives, J.~Smolic and M.~Smolic,
Phys.\ Rev.\ {\bf D73} (2006) 064007,
{\tt hep-th/0509007}.


\bibitem{pirrone}
M.~Pirrone,
JHEP {\bf 0612} (2006) 064, {\tt hep-th/0609173}.

\bibitem{imeroni}
E.~Imeroni and A.~Naqvi,
JHEP {\bf 0703} (2007 034), {\tt hep-th/0612032}.

\bibitem{giant-pp}
K.~Skenderis and M.~Taylor,
JHEP {\bf 0206}, 025 (2002)
{\tt hep-th/0204054}.\hfill\break
H.~Takayanagi and T.~Takayanagi,
JHEP {\bf 0212} (2002) 018,
{\tt hep-th/0209160}.

\bibitem{hamilton}
A.~Hamilton and J.~Murugan, {\em Giant gravitons on deformed
PP--waves},\hfill\break {\tt hep-th/0609135}.


\bibitem{giant-vibrations}
S.R.~Das, A.~Jevicki and S.D.~Mathur,
Phys.\ Rev.\ {\bf D63} (2001) 024013,\hfill\break
{\tt hep-th/0009019}.

\bibitem{wilsonloopTemp}
S.J.~Rey, S.~Theisen and J.T.~Yee,
Nucl.\ Phys.\ {\bf B527} (1998) 171,
{\tt hep-th/9803135}. \hfill\break
A.~Brandhuber, N.~Itzhaki, J.~Sonnenschein and S.~Yankielowicz,
Phys.\ Lett.\ {\bf B434} (1998) 36,
{\tt hep-th/9803137}
and
JHEP {\bf 9806} (1998) 001,
{\tt hep-th/9803263}.

\bibitem{bs}
A.~Brandhuber and K.~Sfetsos,
Adv. Theor. Math. Phys.  {\bf 3} (1999) 851,\hfill\break {\tt hep-th/9906201}.

\bibitem{tipologio}
I.S.~Gradshteyn and I.M.~Ryzhik,
{\it Table of integrals, series and products}, fifth edition
(Academic Press, New York, 1994).

\bibitem{BF}
P.~Byrd and M.~Friedman, {\it Handbook of Elliptic Integrals for
Engineers and Physicists}, second edition, (Springer Verlag,
Heidelberg, 1971).

\bibitem{Callen}
H.B. Callen, {\it Thermodynamics and introduction to
thermostatistics}, 2nd edition, (John Wiley \& Sons, New York,
1985).

\bibitem{ASSiampos}
S.D.~Avramis, K.~Sfetsos and K.~Siampos,
Nucl. Phys. {\bf B769} (2007) 44,\hfill\break
{\tt hep-th/0612139}.

\bibitem{ASSi2}
S.D.~Avramis, K.~Sfetsos and K.~Siampos, {\em Stability of string
configurations dual to quarkonium states in AdS/CFT}, {\tt
0706.2655 [hep-th]}.

\bibitem{buscher} T.~Buscher,
Phys. Lett. {\bf B194} (1987) 59 and Phys. Lett. {\bf B201} (1988)
466.

\bibitem{bho} E.~Bergshoeff, C.M.~Hull, T.~Ort\'{\i}n,
Nucl. Phys. {\bf B451} (1995) 547, {\tt hep-th/9504081}.

\end{thebibliography}
\end{document}